\documentclass{sigma}
 
\begin{document}



\ShortArticleName{}

\ArticleName{Discretising differential geometry via a new product on the space of chains}
\title{Discretising differential geometry via a new product on the space of chains}
\Author{Vivien de Beauc\'e~$^\dag$ and Siddhartha Sen~$^\ddag$}
\AuthorNameForHeading{V. de Beauc\'e and S. Sen}

\Address{$^\dag$~Department of Physics, Hokkaido University, Sapporo, 060-0810 - Japan } 
\EmailD{debeauce@particle.sci.hokudai.ac.jp} 
\Address{$^\ddag$~I.A.C.S Jadavpur, Kolkatta 700032 - India}
\Address{$^\ddag$~School of Mathematical Sciences, UCD, Belfield, Dublin 4, Ireland}
\EmailD{sen@maths.ucd.ie} 


\Abstract{A discretisation of differential geometry using the Whitney forms of 
algebraic topology is consistently extended via the introduction of a 
pairing on the space of chains. This pairing of chains enables us to give a definition 
of the discrete interior product and thus provides a solution to a notorious puzzle in discretisation 
techniques. Further prescriptions are made to introduce metric data, as a 
discrete substitute for the continuum vielbein, or Cartan formulation. 
The original topological data of the de Rham complex is then recovered 
as a discrete version of the Pontryagin class, a sketch of a few examples of the technique is also
provided.  A map of discrete differential geometry into the non-commutative geometry of graphs is 
constructed which shows in a precise way the difference between them. }




\vspace{10mm} 




\newcounter{within}
\newcounter{within3}
\newcounter{within2}
\newcounter{within4}
\newtheorem{theo}{Theorem}[within]
\newtheorem{examp}{Example}[within2]
\newtheorem{lemme}{Lemma}[within3]
\newtheorem{defi}{Definition}[within4]



      


In a wide range of problems in physics and engineering, computational 
challenges have become a major hurdle, for instance in fluid dynamics 
\cite{ricca}, electromagnetism \cite{bossa},  and in fundamental physics.

Many of the issues may be formulated as algebraic problems: how can 
we impose the properties of the continuum theory in the algorithm, in 
the definition of fields, or impose constraints on the evolution of the 
system, if we have in mind a symmetry of the system such as rotational 
symmetry. As is known, the most concise way to formulate topological and 
geometrical properties is by use of the coordinate free language of 
differential forms to represent the fields and use the mathematical 
operations on them. The latter are namely the wedge product, the exterior 
derivative, the Hodge star and finally the interior product (or contraction 
operation), respectively denoted by the symbols $(\wedge, \, d,\, \star, 
\, i_X)$ where $X$ stands for a vector field. With this collection of 
operations we can define all operators of interest, such as the Lie and 
covariant derivatives. 

To define such rules on a lattice, or more precisely on a simplicial 
or hypercubic complex is the problem we address in this article. 
Much work has been done in this area, for instance by discretisation of 
the underlying topological theory \cite{warner, hiptmair, dodzuik, albevario, GD, 
starproduct, scott} and introducing relevant 
operators such as the discrete analogue of the contraction operator \cite{bossa, hirani, desb}, 
even though some algebraic properties are lost in the process. This structure is often referred to as 
a discrete exterior calculus. Of course, which properties are relevant 
will depend on the application at hand, for example in the 
electromagnetic theory, the duality between the electric and the magnetic fields is 
important. To find the optimal discretisation from an algebraic point 
of view is the problem of interest here. 

We will develop further our earlier proposal \cite{deBeauce:2004cw}. In 
Section \ref{thescheme}, we define an extension of the Whitney map, which maps discrete
$p$-dimensional objects ($p$-chains) to $p$-forms, to $\bar{W}$, which are maps from the Cartesian product 
of chains to forms, in such a way that the interior product is well-defined and exact 
\footnote{in the sense that from the continuum viewpoint, the result of 
the operation is correct. Not to be confused with exact forms of algebraic 
topology}. A key construction is to use the map $\bar{W}$ introduced to give an alternate definition of two 
of the basic operations of differential geometry, namely the discrete wedge and Hodge star. 
In Section \ref{further}, we will propose a new approach to discretising the metric in the same 
spirit although additional prescriptions are required, in Section 
\ref{relation}, a map between our discrete differential geometry and Non-commutative geometry of graphs is constructed
this clearly shows the different between these two structures. The examples provided here are for two dimensions.
In Section 5, we discuss the applications we are currently pursuing, namely, the Reidemeister torsion calculation in topological field theory, 2D lattice gravity and the discretisation of Maxwell's equations.
\newpage

\tableofcontents
\section{Introduction}
\label{intro}
\label{sec}
\stepcounter{within}
\stepcounter{within2}
\stepcounter{within3}
\stepcounter{within4}
\subsection{Importance of the Homology ring}
\label{una}
First, we recall some basic concepts of topology. The space of chains 
\begin{equation}
C_{\star}(K;\, \mathbb{R}) \doteq \bigoplus_{p} C_{(p)}(K;\, 
\mathbb{R})
\end{equation}
contain linear combinations of simplices with coefficients in the field 
$\mathbb{R}$, on which one defines a collection of mappings, namely the 
co-chain space $C^{\star}(K;\; \mathbb{R})$ which in turn, within the 
de Rham co-homology setting, may be identified with a finite subspace of 
the space of differential forms, called Whitney forms \cite{Whitney} 
(see the map $W$ below). 

To our knowledge, all approaches currently considered to the discrete 
exterior calculus take all mappings from chains onto chains. To be 
more precise, the operations are all surjective in the space of chains, 
e.g, the discrete wedge product is defined as 
\begin{equation}
\wedge^K:\; C_{(p)}(K;\, \mathbb{R}) \times C_{(q)}(K;\, \mathbb{R}) 
\longrightarrow C_{(p+q)}(K; \,\mathbb{R}).
\end{equation}
 The surjectivity in the definition of the discrete maps, we will think 
of as an assumption referred to as {\bf A\bf}. One can also appeal to the simplicial approximation theorem to motivate the latter. 

Because they were based on homological operations, for example, the dual complex 
$L$ is constructed in order to define the discrete Hodge star (there 
are two Hodge star maps $\star^K$ and $\star^L$ mapping the complex $K$ 
to its dual $L$ and back), the cup product appears in the definition 
of the discrete wedge product; such mappings respect the homology ring \footnote{The cup product as a discrete 
wedge, and the discrete Hodge star satisfy respectively
\begin{eqnarray}
H_{(p)}(K; \, \mathbb{R}) \cup H_{(q)}(K;\, \mathbb{R}) &\cong& 
H_{(p+q)}(K;\, \mathbb{R}), \\
H_{(p)}(K;\, \mathbb{R}) &\cong& H_{(n-p)}(K;\, \mathbb{R}).
\end{eqnarray}
}. 

Let $\sigma_{i, (p)}$ denote the $i$-th $p$-simplex, and $\lambda_{i, (p)}$ 
coefficients in $\mathbb{R}$. With the homology class of the chain 
\begin{equation}
\sigma_{(p)} = \sum_i \lambda_{i, (p)}\sigma_{i, (p)}
\end{equation}
 written as ${\bf [\bf} \sigma_{(p)} {\bf ]\bf}$, then 
\begin{equation}
[\sigma_{(p)}] \wedge^K [\sigma_{(q)}] = [\sigma_{(p+q)}]
\end{equation}
Similarly, using the discrete Hodge star from $K$ onto $L$,
\begin{equation}
\star^K [\sigma_{(p)}^K] = [\sigma_{(n-p)}^L], 
\end{equation}
where $n$ is the dimension of the complex.
The fact that the homology ring is captured does not however mean that 
the operations themselves are exact, not even that they satisfy their 
defining algebraic relations. We will see in the case of the discrete 
wedge that it is non-associative but in a way that is mild.

The virtue of respecting the homology ring is easily understood, the 
operations give a chain that is in the correct homology class. That an operation 
is approximate does not alter this feature. Thus, enforcing this constraint from topology in the construction (which we will call the 
"Homological constraint", {\bf C\bf}), leads to a lack of exactness of the 
operations, apart from the discrete exterior derivative. 

The property of "exactness" denoted {\bf P\bf} below means that for a given simplex, the operation is numerically exact for any simplex as can be checked by application of the map $W$ or $\bar{W}$ as we will see shortly.
\subsection{Lack of exactness of the operations}
In order to understand how this approximation appears concretely, for the sake of the discussion, we enforce both the assumption {\bf A\bf} and the 
constraint {\bf C\bf}. To a given chain $\sigma_{i, (p)}$, we associate a 
differential form $W(\sigma_{i, (p)})$. The list of such forms in the 
case of the hyper-cubic complex can be found in Example \ref{exp}, below. 

The immediate problem is that of exactness: take a Whitney form, say 
$W([01])$, while computing the exterior derivative $dW([01])$ gives a 
linear combination of Whitney forms and the exact result, there is 
no way to obtain closure of the Hodge star on the space of Whitney 
forms, i.e the list of Whitney forms does not include the Hodge duals of 
Whitney forms. In general, given a chain $\sigma$, there is no chain 
$\sigma'$ such that $\star W(\sigma) = W( \sigma')$. Similarly, a 
discrete wedge, also to be defined below can only be approximate. From this 
simple observation, we conclude that a discrete wedge product or Hodge 
star satisfying {\bf A\bf} cannot be exact. 

Nevertheless, such discretisation has remarkable algebraic properties, 
since the discrete exterior derivative $d^K$, which in the simplicial 
setting is the co-boundary operator denoted $d^K$ (which is motivated by the 
Stokes theorem Eq. \ref{stk}) is an exact operation and we do have
\begin{equation}
dW = W d^K.
\end{equation} 
Although the discrete wedge product $\wedge^K$ is 
numerically approximate, the Leibniz rule is satisfied, and similarly, 
although the discrete Hodge star operator is approximate, it satisfies the 
correct algebraic properties and gives an approximate adjoint
\begin{equation} 
\delta^K = (-1)^{n(p+1)+1} \star^L d^L \star^K
\end{equation}
 for the discrete operator $d^K$, which satisfies 
\begin{equation}
(\delta^K)^2 = 0,
\end{equation}
which may also be seen as a consequence of the exactness of $d^K$. The approximation coming from 
the other operation used, namely the discrete Hodge star is done within the correct co-homology class. 

To summarize, it has been found that under the assumption {\bf A\bf}, a 
collection of discrete maps for $\wedge^K$, $d^K$ and $\star^K$ and 
similarly on the dual complex $L$ satisfying the constraint {\bf C\bf} is 
well-defined. Let us now turn to the question of discretising the 
interior product $i_X$.

\subsection{Importance of exactness for the discrete interior product}
When one comes to the problem of supplementing the model with an 
interior product (the contraction operation), it is evident that:
\begin{itemize}
\item
The operation, defined from chains onto chains (i.e satisfying {\bf 
A\bf}) can only be approximate (violating {\bf P\bf}). 
\item
The algebraic properties do not hold, notably the anti-derivation 
property. 
\end{itemize}
The situation is similar to that of the discrete wedge product in that 
an approximation must be made, but in the present case, a discrete 
operator for $i_X$ using the operations $(\wedge^K, d^K, \star^K)$ will not 
satisfy exactness as was the case for $d^K$, thus some algebraic 
properties are spoiled. Such proposal was initiated by Hirani \cite{hirani, desb}, 
who discretised  the following continuum expression for the contraction operation on a $n$-dimensional manifold 
\begin{equation}
\label{hiraf}
i_X \omega^{(k)} = (-1)^{k(n-k)}\star \omega^{(k)} \wedge \star X^b,
\end{equation} 
where $X^b$ is the dual form to the vector field $X$ and $\omega^{(k)}$ is a $k$-form. And so one may 
use this formula directly in the discrete setting with the discrete version of the operations 
$\wedge, \star, d$ in place of the continuum ones.

While the convergence of the operation in the continuum limit has been 
shown, from an algebraic point of view it is incomplete, due to the 
non-associativity of the wedge product alluded to above. One motivation 
for building such operator with more algebraic properties is to consider 
the Lie derivative given by the Cartan formula,
\begin{equation}
L_X = i_X d + d i_X.
\end{equation} 

After defining the usual operations which allow to discuss topology in 
Section \ref{thescheme}, we move on to the central construction, first, 
we introduce a larger space using a product on chains, this requires 
extending the Whitney map to this larger domain, the extension is not 
trivial and we are lead to check that the co-homology still holds. The 
natural one-to-one correspondence between chains and co-chains still 
holds while a larger subspace of differential forms is constructed. After the introduction of the interior product $i_X$, we discuss its 
algebraic properties which are identical to the continuum ones, and derive 
the Lie derivative, arguing in what sense the Jacobi identities are 
piecewise satisfied. Then, following the logic of the introduction of the 
product space, we suggest a new Hodge star (denoted $\bigstar^E$) which 
does not require the introduction of a dual complex (this map was 
introduced in the discussion of lattice QCD in \cite{deBeauce:2005ny}). In 
the latter part, we present the simplicial version of the above, for 
completeness.

In Section \ref{further}, we introduce metric in the present scheme, 
for this purpose, we find that the map $\bar{W}$ is essential. We 
introduce a discrete version of the vielbein and of the Cartan structure equations and then consider the Pontryagin class which leads to 
topological invariants. In Section \ref{relation}, we discuss the relation of 
the present approach with the non-commutative geometry on finite sets 
as a discrete differential geometry proposed by Woronowicz \cite{wor} 
(see also \cite{Dimakis:1994qq}). A relation is established 
after removing the smoothness associated to Whitney forms and to their 
generalization given here. We show how to extract the non-commutative 
setting from the present formulation, the two being not equivalent. A discussion of some applications to physics follows in Section 5.

Before starting the construction, we ask the following question: what 
features of a manifold can we expect to capture on a simplicial complex ?

\subsection{Local versus global, simplices and manifolds}

Consider the triangulation of a $n$-dim manifold $M$ by a simplicial 
complex $K$, in the sense of singular homology it can be identified as a 
combinatorial manifold and as a piecewise linear (PL) manifold, that is we 
have a manifold structure. In the standard simplex 
\begin{equation}
\triangle^n = \{ (x_0,\ldots, x_{n+1}):\; 0 \leq x_i \leq 1, \; \sum_i 
x_i =1 \} \subset \mathbb{R}^{n+1},
\end{equation}
associated to each 
simplex, the continuous mappings $\sigma:\,\triangle^n \longrightarrow K$ are 
the singular simplices of K. Coordinates are thus provided, and it is in these coordinates that we will 
define the basis of Whitney forms and the extension defined here under 
the map $\bar{W}$. 

Given two open sets $U_{\alpha}$ and $U_{\beta}$, associated to two neighbouring top degree simplices $\sigma_{\alpha,\, (n)}$ and $\sigma_{\beta,\, (n)}$, the 
transition function 
\begin{equation}
\phi_{\alpha\beta} = \phi_{\alpha}\, o\, \phi_{\beta}^{-1}
\end{equation}
defined on $U_{\alpha} \cap U_{\beta}$ is a homeomorphism. However it is  not a diffeomorphism and so 
the PL manifold we have constructed is not a differentiable manifold 
since in that case the transition functions should be elevated to a 
diffeomorphism. That the latter cannot be defined here is easy to understand, 
the intersection of two $p$-simplices is a $(p-1)$ simplex, again in 
the singular homology one may introduce the $(p-1)$ standard simplex which has lower dimensionality, therefore the 
transition function does not have the right differentiability, only $(n-1)$ 
partial derivatives are defined.

\begin{figure}[H]
\label{fig1}
{\par\centering \includegraphics{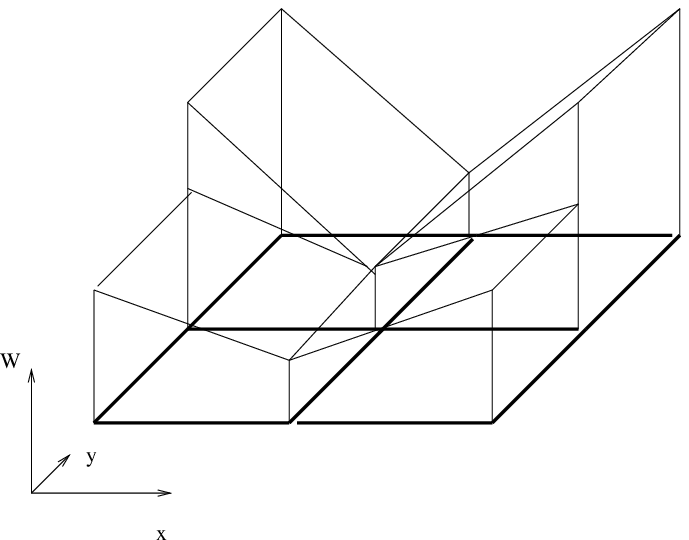} \par}
\caption{\label{fig:with}The two dimensional plane with a square 
complex, coordinates $(x,\, y)$ and a Whitney form with function coefficient 
sketched in the $W$-axis. In FIG. \ref{fig-eg} it corresponds to the co-chain $W(2 [03] + [12] + 3[35] + [24] + 2[67] + 3[78])$}
\end{figure}

That we do not have a differentiable manifold is an important 
limitation when we come to consider global objects, such as those associated to 
vector fields, for example Lie group flows, and topological invariants 
computed not in the Homology (which is trivial using the complex) but 
in the space of forms $\Omega(M;\, \mathbb{R})$, such at the 
Gauss-Bonnet theorem considered in section \ref{further}. A look at the space of 
Whitney forms in 2D (Example \ref{exp}) will reveal that only the zero co-chains are 
continuous, while the one-cochains are not continuous (See FIG.\ref{fig1}) and 
so are the two co-chains. All these are only continuous in the standard 
simplex $\triangle^n$, i.e locally, by which we mean for the given 
$n$-degree simplex $\sigma_{i, (n)}$. 
In not having a smooth structure globally, we are lead to construct 
operations locally where there is a smooth structure, as announced the 
interior product will be exact in the sense of locally exact. 

For example, in 2D, suppose the Euler class is discretised as a Whitney two-form, 
consider the integral of the co-chain, say mapping the two-chain $e(K)$ to the Whitney two form $e(M) \doteq W(e(K))$, 
\begin{equation}
\int_{M}  e(M)\doteq \sum_i \int_{|\sigma_{i, (2)}|} e(\sigma),
\end{equation}
but this is not exact, suppose as in Section \ref{further} we may take the gauge transformation specified by 
$P$ such that in each simplex $P^{-1} \mathcal{R} P$ is flat, then the Euler characteristic is zero.  
We are missing the contribution from distributions over the edges 
where the 2-form changes discontinuously. So one would have for each link 
a delta function\footnote{The use of delta function for the Euler class is an example of a current \cite{grif}} with a weight factor so that $e(M)$ should be 
\begin{equation}
e(M)' = e(M) + \sum_j K_j\delta( x- x_j),
\end{equation}
The factors $K_j$ would be determined by the Gauss-Bonnet theorem, and 
correspond to the subtle play between integration domain and the Euler 
class which is itself a measure, this interplay gives the topological 
invariant in the continuum theory. In the present case one could imagine 
making a manifold which imitates the present simplicial complex 
situation in which the field are now $\mathcal{C}^{\infty}$-functions, in the limit were 
the vielbein gets close to the simplicial field, we recover a delta 
function as above. Having understood this limitation, namely the lack of 
relation between local and global, we will proceed by means of the 
overall normalization $N$ for the Euler class 
\begin{equation}
e(M) = N \sum_i e(\sigma_{i,\, (2)})
\end{equation} 
as will  be described in section \ref{further}. This seems a round 
about way to calculate the Euler characteristic which is $\chi(K) = V-E+F$, or using 
the homology groups which are readily calculable in the chain space 
using the boundary operator $\partial^K$. 

As seen in FIG. 1, the type of vector field satisfying {\bf 
A\bf} introduced is not smooth, for example their divergence
\begin{equation}
\nabla . X 
\end{equation}
is ill defined over a boundary unless the vector is constant globally 
because of the hopping of the vector field across cells. Also, if the vector field is represented as its dual (in the sense of metric) one chain, its 
divergence is zero in each simplex. It is then clear that the flow of a vector 
field can only be defined locally, it is approximate, but 
nevertheless the formula for the Lie derivative is exact within the cell and 
compatible with the de Rham cohomology.

Our aim is to approximate the continuum theory of differential geometry, to that effect we construct operations which are locally exact in the sense explained above. On the other hand 
the de Rham co-homology is maintained in the construction, which means 
that the co-boundary operator $d^K$ which acts as a linear operator on 
the chain space, is compatible with the contraction for each simplex. In 
Table \ref{T:deux} we take schemes that may be considered with their list of operations and we 
give on the right hand side the properties satisfied by them as we will see shortly. 
\begin{table}[here]
\begin{center}
\begin{tabular}{|c|c|c|}
\hline
Ring of operators  &  Properties \\
\hline
& \\
$\{ d^K,\, i_X,\, \bigstar,\, \bigwedge \}$ & \text{{\bf A,\, C,\, 
P\bf}} \\
& \\
\hline
& \\
$\{d^K,\, i_X,\, \bigstar,\, \wedge^K \}$ & \text{{\bf C\,  \bf}} \\
$\{d^K,\, i_X,\, \star, \bigwedge\}$ & \\
& \\
\hline
& \\
$\{d^{K,L}, \wedge^{K,L}, \star^{K,L}\}$ & \text{{\bf A,\, C \bf}}\\
& \\
\hline
\end{tabular}
\end{center}
\caption{\label{T:deux}The various schemes and their properties}
\end{table}
This introduction concludes with the list of the new basic operations and where they are defined in the text: \\

\noindent
$\times$, Cartesian product, Eq. \ref{bar}, \\
$\bar{W}$, Extended Whitney map, Eq. \ref{wbar}, \\
$\varphi^{(0)}$, Map to function coefficient of a form, Eq. \ref{varphi0}, \\
$D^K$, Extension of $d^K$ to the product space, Eq. \ref{dkdef}, \\
$\bigstar^E$, Discrete Hodge star on the product space, Eq. \ref{stare}, \\
$\bigwedge$, Discrete wedge on the product space, Eq. \ref{bigwedge}, \\
$\bigstar^{\theta}$, Curved metric Hodge star, Eq. \ref{hod}, \\
$\rho$, Morphism from non-commutative form to usual differential form, Eq. \ref{rho}, \\
$T$, Truncation map, Eq. \ref{trunca}.

\section{The scheme for discrete differential geometry}
\label{thescheme}
\stepcounter{within}
\stepcounter{within2}
\stepcounter{within3}
\stepcounter{within4}
\subsection{Topology}
\subsubsection*{The mappings between chains and co-chains, $W$ and $A$}
Consider a hypercubic complex $K$, for the moment we will work with a 
square complex. A $p$-chain is a linear combination of hypercubes,
\begin{equation}
\sigma_{(p)} = \sum_{i} \lambda_{i, (p)} \sigma_{i , (p)},
\end{equation}
the associated co-chain $\sigma^{(p)} = W(\sigma_{(p)})$ is represented 
as a differential form, it has well-defined value in the standard 
simplex associated to each $n$-hypercube of which $\sigma_{i, (p)}$ is a 
face. One associates to the standard simplex the Cartesian coordinate 
system 
\begin{equation}
| \sigma_{i, (p)} | \doteq [0,1]^p.
\end{equation}
To construct the co-chain as a differential form, the Whitney map and 
the de Rham co-homology inner product are used, the following can be seen 
as part of the definition of the map $W$, 
\begin{equation}
(\sigma_{i, (p)}, W(\sigma_{j, (q)})) \doteq \int_{| \sigma_{i, (p)}|} 
W( \sigma_{j, (q)}) = \delta_{i,j} \delta_{p,q}.
\end{equation}
The Whitney map is linear, and since $d^K$ is the co-boundary operator 
on the chain space, then
\begin{equation}
\label{dw}
d W = W d^K.
\end{equation}
We note that the operation $d^K$ is exact in the interior of every 
hypercube, this follows directly from Eq. \ref{dw}. This statement may be 
expressed as a commutative diagram illustrating the fact that $W$ is a 
functor from the space of chains to the space of co-chains, mapping the 
morphism $d^K$ to $d$. We will refer to this property of exactness as 
{\bf P\bf}.
\begin{examp} Whitney forms for the square complex \\
\label{exp} 

Whitney forms for a 2D square complex \cite{samik}. In each square 
choose a vertex as the origin, and choose an orientation, then introduce 
the coordinates $(x,y)\in [0,1]^2$. Given a square, the chain $[0123]$ 
let $[0]$ be the origin, $[1]$ with coordinates $(1,0)$, $[2]$ with coordinates 
$(1,1)$ and $[3]$ as $(0,1)$. Then, the Whitney zero forms are:
\begin{equation}
\label{zerof}
W([0]) = (1-x)(1-y), \; W([1]) = x (1-y), \; W([2]) = xy, W([3]) = 
(1-x)y,
\end{equation}
the Whitney one forms:
\begin{equation}
W([01]) = (1-y)dx, \, W([12]) = x dy, W([32]) = ydx, \, W([03]) = 
(1-x)dy,
\end{equation}
and the only Whitney two form is
\begin{equation}
 W([0123]) = dx \wedge dy.
\end{equation}
This is the complete list of co-chains as differential forms, in the 
chart with coordinates $(x,y)$ corresponding to the square $[0123]$, see 
FIG. \ref{fig-eg}.

When expressing the Whitney form in coordinates, it will be convenient 
to indicate in which square it is defined, we will do that by putting 
the unit function in front of the expression like 
\begin{equation}
W([32]) = 1_{[0123]} y dx + 1_{[3245]} (1-y) dx,
\end{equation}
indicating the two squares in FIG. \ref{fig-eg} (top degree simplices in which $W([01])$ has 
non-zero value). This last expression being the global definition of 
$W([01])$.
\end{examp}

The de Rham map $A$, is the left inverse of the Whitney map ($AW = 1$), 
it is defined over the space of forms, like integration itself, the 
co-chain space being seen as a subspace of the space of forms. For $\omega^{(p)} 
\in \Omega^{(p)}(M;\, \mathbb{R})$, define
\begin{equation}
A ( \omega^{(p)} ) \doteq \sum_i \sigma_{i, (p)} \int_{| \sigma_{i , 
(p)} |} \omega^{(p)}.
\end{equation}
It is immediate that Stokes' theorem may be expressed as
\begin{equation}
\label{stk}
A d = d^K A.
\end{equation}
The property of exactness is lost as soon as one introduces the 
discrete wedge product $\wedge^K$. 
\subsubsection*{The discrete wedge product}
The various mappings are in place now to introduce the discrete wedge 
product $\wedge^K$, it is combinatorial in the sense that it involves 
the continuum wedge product:
\begin{equation}
\label{dwedge}
\sigma_{i,(p)} \wedge^K \sigma_{(j, (q)} \doteq  N(p,q) A(W(\sigma_{i, 
(p)}) \wedge W(\sigma_{j, (q)})),
\end{equation}
where the numerical factor $N(p,q)$ is a normalization \footnote{For 
the square complex it is $N(0,0) = N(0,1) = N(1,0) = 2, N(1,1) = N(2,0) = 
N(0,2) = 4$. Note that $N(p,q) = N(p+q)$, so that the normalization is 
well-defined.}. It follows immediately that the operation satisfies the Leibniz rule, 
expressed as
\begin{equation}
d^K (\sigma_{i, (p)} \wedge^K \sigma_{j, (q)}) = d^K \sigma_{i, (p)} 
\wedge^K \sigma_{j, (q)} + (-1)^p (\sigma_{i, (p)}) \wedge^K d^K 
\sigma_{j, (q)}.
\end{equation}
and is easily proved.
The discrete wedge product for the hypercubic complex is better behaved 
than the simplicial one; both are strictly speaking non-associative, 
but the hypercubic one is almost associative in the sense that one may 
permute the three chains $\sigma_{i, (p)}, \; \sigma_{j, (q)},\; 
\sigma_{k, (r)}$, taking care of the signs in such a way that the wedge is 
associative. So we may write symbolically
\begin{equation}
\sigma_{i, (p)}\wedge^K \sigma_{j, (q)} \wedge^K \sigma_{k, (r)} \doteq 
 \max_{\mathcal{P}_{(ijk)}} (-1)^{\mathcal{P}}\sigma_{i, 
(p)}\wedge^K \sigma_{j, (q)} \wedge^K \sigma_{k, (r)}.
\end{equation}
where $\mathcal{P}$ specifies the sign resulting from permuting the chains.
The reason for the almost associative nature is that in some cases, the 
result might be zero, hence the reshuffling. 
\begin{examp} Discrete wedge product \\

Label the vertices of a 2D hypercubic complex by a pair of integers 
$i$ and $j$, one as integer $X$ coordinate and $Y$ coordinates (not to be 
confused with standard simplex coordinates). We find that
\begin{equation}
([(i,j)] \wedge^K [(i+1,j)]) \wedge^K [(i,j) (i+1, j)] = 0,
\end{equation}
while 
\begin{equation}
[(i,j)] \wedge^K ([(i+1,j)] \wedge^K [(i,j) (i+1, j)]) = [(i,j)(i+1,j)] 
\end{equation}
is the desired result, hence the need for reshuffling before inserting the 
brackets. 
\end{examp}
The operation of reshuffling up to sign is an anti-automorphism in the 
space of forms, it does not modify the co-homology. 
\subsubsection*{Discrete Hodge star $\star^K, \, \star^L$} 
The last operation to be discussed in this introduction is the Hodge 
star operator. The procedure is to introduce a dual complex $L$, and the 
subdivided complex $B$ which consists of all hypercubes arising from 
superimposing $L$ onto $K$. The Hodge star operator is defined as a map 
between the two complexes (see for example \cite{Us} for more details), 
\begin{eqnarray}
\star^{K}&:& C_{(p)}(K;\; \mathbb{R}) \longrightarrow C_{(n-p)}(L;\; 
\mathbb{R}) \\ 
\star^{L}&:& C_{(p)}(L;\; \mathbb{R}) \longrightarrow C_{(n-p)}(K;\; 
\mathbb{R}),
\end{eqnarray} 
after introducing a Whitney map for $L$ as well as for $K$, we 
introduce the scalar product for the Euclidean manifold, defining the norm,
\begin{equation}
(\sigma^K_{i, (p)} , \star^{K} \sigma^K_{j, (q)}) \doteq \int_{M} 
W^{BK} ( \sigma^K_{i, (p)} ) \wedge W^{BL}( \star^{K} \sigma^K_{j, (q)}),
\end{equation}
where $W^{BK}$ and $W^{BL}$ are the expression for the Whitney map on the finer space $B$.
One finds that the adjoint 
\begin{equation}
\delta^K = (-1)^{n(p+1)+1}\star^L d^L \star^K,
\end{equation}
is simply the boundary operator $\partial^K$, up to sign (see for example 
\cite{GD}). The Hodge star thus defined satisfies the defining properties
\begin{eqnarray}
\delta^K &=& (-1)^{n(p+1) +1}\star^L d^L \star^K, \\
\star^K \star^L &=& (-1)^{np+1},
\end{eqnarray}
and respectively interchanging $L$ and $K$ suffixes.
Note that the discrete Hodge star is not exact. Looking back at Example 
\ref{exp}, we see that $W([0])$ is as given in the list $W([0])= (1-y) 
(1-x)$, while $\star W ([0]) = (1-x)(1-y) dx\wedge dy$ is not in the 
list, we discussed this fact already in the introductory part.
\subsection{Geometry}
A product on chains is now introduced, it is motivated by the map $\bar{W}$ used in the discrete 
interior product defined subsequently. The product $\times$ will 
provide us with an exact discrete interior product, and we will use it in a 
way that respects the co-homology.

\begin{defi}{Product $\times$ and extended Whitney map $\bar{W}$} \\
 
A product denoted $\times$ is introduced on the space of chains 
\cite{deBeauce:2004cw}. It is a Cartesian product, the ordered pairing 
\begin{equation}
\label{bar}
\sigma_{i, (p)} \times \sigma_{j, (q)} = (\sigma_{i, (p)}, \sigma_{j, 
(q)}). 
\end{equation}
A map 
\begin{equation}
\label{wbar}
\bar{W}: C_{(p)} (K; \, \mathbb{R}) \times C_{(q)}(K;\, \mathbb{R}) 
\longrightarrow \Omega^{(q)}(M;\, \mathbb{R})
\end{equation}
 is introduced, defined as  
\begin{equation}
\label{varphi0}
\bar{W} ( \sigma_{i, (p)} \times \sigma_{j (q)}) \doteq \varphi^{(0)} ( 
W (\sigma_{i, (p)}) )\wedge W( \sigma_{j, (q)} )
\end{equation}
where $\varphi^{(0)}$ maps a Whitney form to its function coefficient 
in the standard simplex coordinates.
\end{defi}

\begin{examp} Generic square lattice. \\

For a 2D square complex $K$, see FIG. \ref{fig-eg} the chain 
corresponding to a given vertex, say vertex $[2]$ which is the chain $\sigma_{2, 
(0)}$, gives the form $W(\sigma_{2, (0)})$, it has non-zero value over 
the four squares bounding it. 

The edge $[32]$, parallel to the $x$-axis in the embedding 
$\mathbb{R}^2$ gives $W([32])$ with non-zero value over the two squares of which it 
is a bounding edge. Finally, the square $[0123]$ gives a differential 
form which has non-zero value in that square.
\end{examp}
 This limitation of the non-zero support, is one feature of the product 
$\times$, under the map $\bar{W}$, we note that the two co-chains must have some overlap where they are 
both non-zero valued. Such considerations will be of crucial importance 
later, we will have to check that the pair of chains appearing in a 
product are such that the resulting form arising under application of 
$\bar{W}$ is indeed differentiable, e.g if one considers two edges which only 
have non-zero value over edges, not squares, then the differentiability 
does no longer holds. \\

\begin{lemme} Generic constant forms \\

Introduce the generic constant forms, which can be expressed as a sum 
of Whitney forms,
\begin{eqnarray} 
\sigma^0 &\doteq& \sum_{(i,j)} [(i,j)], \\
\sigma^1 &\doteq& \sum_{(i,j)} [(i,j) ( i+1,j)],\; \sigma^2 \doteq 
\sum_{(i,j)} [(i,j) ( i, j+1)], \\
 \sigma^{12} &\doteq& \sum_{(i,j)} [(i,j) (i+1,j) (i+1, j+1) (i , j+1)]. 
\end{eqnarray}
They satisfy $W(\sigma^0) = 1 $, $W(\sigma^1) = dx$, $W(\sigma^2) = dy$ 
and $W(\sigma^{12}) = dx \wedge dy$.

\end{lemme}
These forms are globally constant in the sense that within each top degree simplex they are constant.

We note the relation of the extended Whitney map $\bar{W}$ to the 
original Whitney map $W$. The following lemma follows immediately
\begin{lemme}{Extended Whitney map $\bar{W}$} \\

\noindent
1. If the chain $\sigma_{(p)}$ is such that $W(\sigma_{(p)})$ is a 
generic constant form, that is one of the preceding equations, then
\begin{equation}
\bar{W} ( \sigma_{(p)} \times \sigma_{i, (q)}) = W( \sigma_{i, (q)}),
\end{equation}
for any $\sigma_{i,\, (p)}$. \\

\noindent
2. The map $\bar{W}$ acts as the Whitney map $W$ on the chain space
\begin{equation}
\bar{W}( \sigma_{i, (p)}) \doteq W( \sigma_{i, (p)} ).
\end{equation}
\end{lemme}
Here, vector fields live in a Euclidean space. So the dual one form 
$X^b$ to a vector field $X$ with respect to the Euclidean metric is 
represented as a one-chain:
\begin{equation}
X^K \doteq A ( X^b).
\end{equation}
The introduction of the discrete vector field is straightforward by the 
very fact that topology has imposed on us a Euclidean metric. Below, we 
will see how curved metric data is incorporated. Using the chain space to 
represent the vector field as opposed to the dual complex is we will find 
natural.

Before introducing the interior product, we need one more map, 
\begin{defi} Discrete contraction on chains (hypercubic complex) \\

Given an edge $\sigma_{i, (1)}$ and a $p$-hypercube $\sigma_{j, (p)}$ both faces of a top degree simplex $\sigma_{k, (n)}$, 
let
\begin{equation}
\Lambda (\sigma_{i, (1)}, \sigma_{j, (p)} ) \doteq \{ \sigma_{k, 
(p-1)}\, |\, \sigma_{k, (p-1)} < \sigma_{k, (n)}, \sigma_{k, (p-1)} \wedge^K \sigma_{j, (p)} =0,\, \sigma_{k, (p-1)} \wedge^K \sigma_{i, (1)} \neq 0 \},
\end{equation}
then,
\begin{equation}
\eta_{\sigma_{i, (1)}} (\sigma_{j, (p)} )\doteq \sum_{\Lambda} 
\sigma_{k, (p-1)}.
\end{equation}
\end{defi}
If one draws the hypercube $\sigma_{j, (p)}$ and an edge $\sigma_{i, 
(1)}$ such that $\sigma_{i, (1)} < \sigma_{j, (p)}$. Then 
$\eta_{\sigma_{i, (1)}} (\sigma_{j, (p)} )$ is simply the sum of all $(p-1)$ faces of 
$\sigma_{j, (p)}$ orthogonal to $\sigma_{i, (1)}$.
\begin{defi} Discrete interior product \\

Given a one chain $X^K$, and a $p$-chain $\sigma_{(p)}$, the discrete 
interior product is defined as 
\begin{equation}
i: C_{(1)}(K;\, \mathbb{R}) \times C_{(p)}(K;\, \mathbb{R}) 
\longrightarrow C_{(1)}(K;\, \mathbb{R}) \times C_{(p-1)}(K;\, \mathbb{R}),
\end{equation}
where
\begin{equation}
i_{X^K} \sigma_{(p)} \doteq X^K \times \eta_{X^K} (\sigma_{(p)}).
\end{equation}
\end{defi}
\begin{examp} Interior product \\

Start with a $1$D example. Consider the edge $[01]$ on the line. Then, 
the Whitney elements are
\begin{equation}
\{ W([0]) = 1-x, \, W([1]) = x , \, W([01]) = dx\},
\end{equation}
and
\begin{equation}
i_{[01]} [01] = [01] \times ([0] + [1]). 
\end{equation}
That is, we took the sum of the two vertices that remain when we remove 
the edge $[01]$. Also, the construction is guided by the embedding in 
the continuum:
\begin{equation}
\bar{W}(i_{[01]} [01]) = \varphi^{(0)}(W([01])) \wedge W([0] + [1]) = 
1. 
\end{equation}
The one chain $[01]$ plays the role of a function 
coefficient so, in this case it is $1$ and hence the result. If you 
contract $[0]$ or $[1]$ with the edge $[01]$, you get zero because the edge 
$[01]$ is not a face of either vertices. 

Consider the 2D example of the square (FIG.~\ref{fig-eg}).
\begin{figure}[h]
         \centerline{
           \scalebox{0.6}{
             \includegraphics{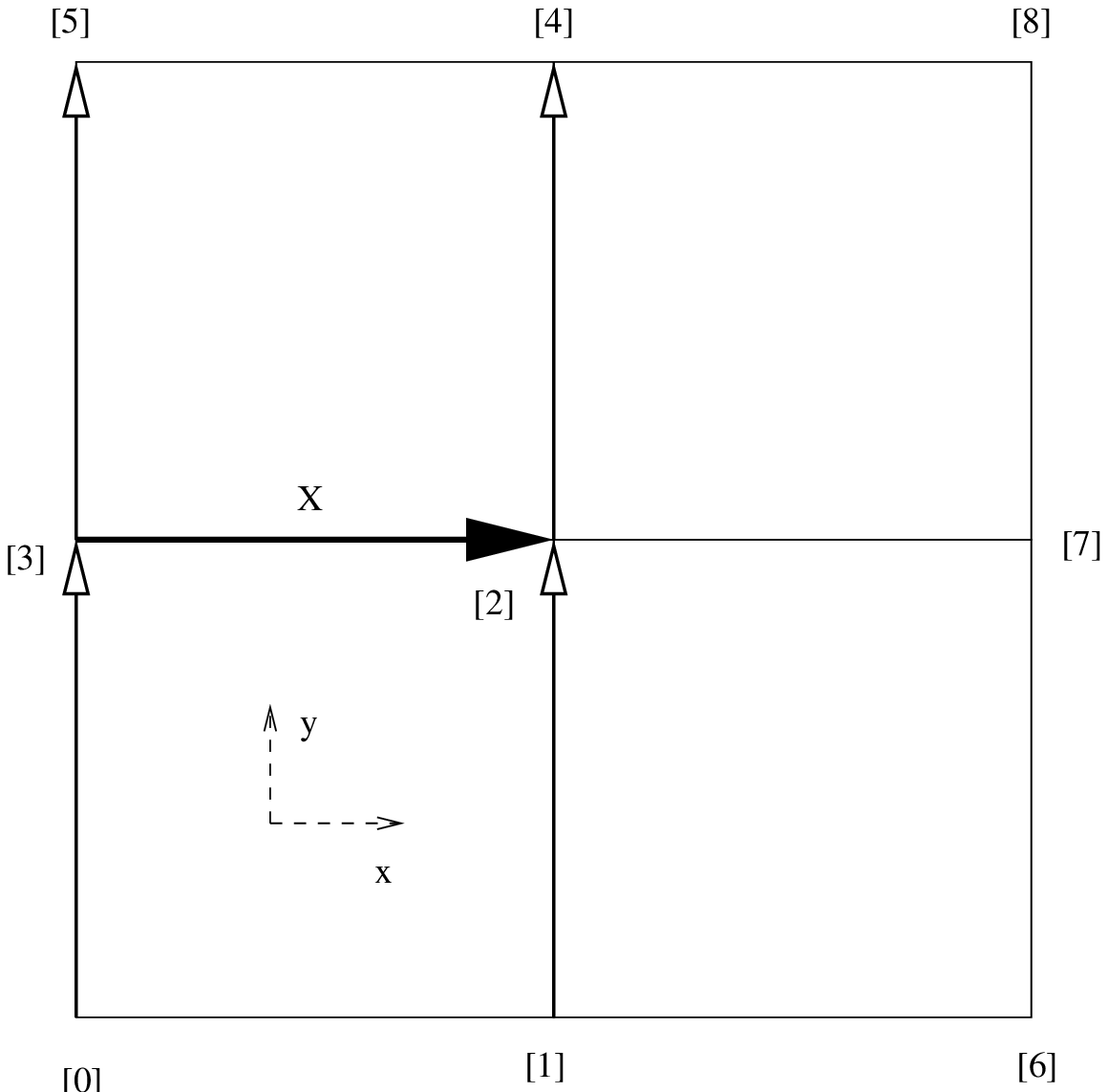}
             }
           }
         \caption{Part of a square complex, with the one-chain $X$ corresponding to the vector field}
         \label{fig-eg}
       \end{figure}
This time there are many combinations of two chains which lead to a 
non-zero contraction, for example the contraction. Let us see three examples of $i_X$
\begin{align}
\bar{W} (i_{[32]} [0123]) &= \bar{W}([32] \times ([03]+[12])) = 
\varphi^{(0)}(y dx) \wedge dy = 1_{[0123]}ydy, \\
\bar{W}(i_{[32]} [32])  &= \bar{W}([32] \times ([3] + [2])) = 
1_{[0123]}y^2 + 1_{[3245]}(1-y)^2 \\
\bar{W}(i_{[32]} [01])  &= \bar{W}([32] \times ([0] + [1])) = 
1_{[0123]}(1-y)y.
\end{align}
The second line shows that the form obtained after 
contraction has non-zero value only in the two squares $[0123]$ and 
$[3245]$.

The first calculation shows how when contracting a $2$-chain, 
we get a linear combination of the edges perpendicular to the edge 
which specifies the vector field. This generalizes obviously to higher 
degree chains: we get a linear combination of the $(p-1)$-simplices which 
are orthogonal to the vector field (an edge). The third line shows that 
although $[32]$ is not a face of $[01]$, the contraction still applies 
because they are parallel within the same cell of highest degree which 
plays the role of the local open set.
\end{examp}
Application of $d^K$ after contraction is prescribed as follows, the 
extension of $d^K$ is defined to be $D^K$ and acts as
\begin{equation}
\label{dkdef}
D^K (i_{X^K} \sigma_{i, (p)}) \doteq 2 X^K \times d^K (\eta_{X^K} 
\sigma_{i, (p)}). 
\end{equation}
\begin{lemme} Compatibility of $i_X$ with the de Rham co-homology \\
\label{lmm}
 
The discrete interior product is compatible with the de Rham cohomology 
in the sense that:
\begin{equation}
\label{div}
\bar{W} D^K i_{X^K} = d \bar{W} i_{X^K},
\end{equation}
\end{lemme}
To prove the lemma, we inspect the list of Whitney elements for the square Eq.\, 
\eqref{zerof} any vertex $[i]$ has within a given top dimensional simplex 
$\triangle^n$ of which it is a face the following expression:
\begin{equation}
W ([i] ) = \Pi_k f^{sgn(\mu (k,i))}_{\mu(k,i)},
\end{equation}
where each factor is either
\begin{align}
f_j &= f^{+}_j \doteq (1-x_j), \\
&or \notag \\
&= f^{-}_j \doteq  x_j. 
\end{align}
Assuming that there are $p$ factors of the type $f^+$ and $(n-r)-(p+1)$ factors of the $f^-$ type we get
\begin{equation}
W( \sigma_{(r)} ) = \Pi_{i=1}^{p} f_{\mu(i,r)}^{-} \Pi_{j=p+1}^{n-r} f_{\mu(j,r)}^{+} \, 
\bigwedge_{k=1}^{r} dx^{\mu_k}.
\end{equation}
Then, starting with the left hand side of Eq.~\ref{div}, we get
\begin{equation}
\bar{W} ( 2 X^K \times \, D^K \eta(\sigma)) = d (\varphi^{(0)}((W (X^K))) \wedge  W (\eta_{X^K}(\sigma))),
\end{equation}
By moving the derivatives 
$\frac{\partial}{\partial x^l}$ across to the left, we pick up all the terms since the vector 
has at least every $f_i$ factor that $\sigma$ has. 

Before we proceed we need to make an additional prescription for the 
wedge product. 
\begin{defi} Discrete wedge and Hodge star on the product space \\

It is defined as follows
\begin{equation}
i_X \sigma_{i (p)} \wedge^K i_Y \sigma_{j (q)} \doteq X \otimes Y 
\times (\eta_{X} \sigma_{i (p)} \wedge^K \eta_{Y} \sigma_{j (q)})
\end{equation}
where $\otimes$ is the tensor product. The above formula generalizes 
easily to the case where there are many terms contracted and then wedged. 

A similar property applies to the discrete Hodge star:
\begin{equation}
\star^K(\sigma_{i (p)} \times \sigma_{j (q)}) \doteq \sigma_{i (p)} 
\times \star^K \sigma_{j (q)}.
\end{equation}
\end{defi}
\begin{theo}
The discrete interior product is exact
\begin{equation}
\bar{W}(i_{X^K} \sigma_{(p)}) = i_{W (X^K)} W (\sigma_{(p)}).
\end{equation}
Furthermore, it satisfies
\begin{equation}
(i_{X^K})^2 = 0,
\end{equation}
and it is an anti-derivation
\begin{equation}
i_{X^K} (\sigma_{i (p)} \wedge^K \sigma_{j (r)}) = (i_{X^K} \sigma_{i 
(p)} )\wedge^K \sigma_{j (r)} + (-1)^p \sigma_{i (p)} \wedge^K i_{X^K} 
\sigma_{j (r)}.
\end{equation}
\end{theo}
At this point, we have established the algebraic tools which allow to 
construct the Lie derivative which we take to be
\begin{equation}
L_{X^K} \doteq i_{X^K} D^K + D^K i_{X^K}.
\end{equation}
Furthermore, the Lie derivative is exact, i.e 
\begin{equation}
\bar{W} L_{X^K} \sigma_{i, (r)} = L_{W X^K} W (\sigma_{i, (r)}).
\end{equation}
This immediately implies that one side of the Jacobi identities can be 
evaluated, namely (by composition of the exact mappings)
\begin{equation}
\label{j}
[L_X, \, L_Y].
\end{equation}
To define the Lie bracket, well, it should be exact in order to match 
Eq. \ref{j}. We note that the Jacobi identities for the Lie
\begin{equation}
\label{jac}
L_{[X,Y]} = [L_X,\, L_Y]
\end{equation}
hold if and only if the Lie bracket is exact, that is if it closes on the chain 
space
\begin{equation}
\bar{W} ([X^K  \,,\, Y^K  ]^K) = [W(X^K), \, W(Y^K)].
\end{equation}

However, the bracket may be expressed in terms of the product space. 
The following can be seen as a definition of the bracket:
\begin{equation}
([X^K  \,,\, Y^K  ]^K) = Z^K,
\end{equation}
where $Z^K$ is such that
\begin{equation}
\bar{W} Z^K = [W(X^K), \, W(Y^K)].
\end{equation}

Note that the introduction of the product $\times$, enforces the role 
of the local open set associated to the top dimensional simplex, so the various terms arising from the evaluation 
of Eq. \ref{jac} have interpretation in each and every open set associated 
to a square.

To think about flows, $\sigma_X$ the flow of $X$. It is well-defined 
per-cell. However, the vector changes abruptly across cells, if $X$ has 
an orthogonal component to a boundary, it changes discontinuously 
across that boundary. 

\subsection{Cohomology}
Before discussing the de Rham co-homology in the context of the product 
space, we must understand how this question arises in the first place. 
In the picture of co-chains as differential forms, as provided by the 
Whitney $W$ and de Rham $A$ maps, the co-chain have global definition. A 
priori, the product space may contradict this crucial feature since the 
product of two chains $\sigma_{i,\, (p)} \times \sigma_{j,\, (q)}$ has 
non-zero value where both cochains have overlapping non-zero value.
\begin{lemme} Co-homology and differentiability \\
Given a pair of chains $(\sigma_{i (p)}, \; \sigma_{j, (q)})$, their 
product $\sigma_{i , (p)} \times \sigma_{j, (q)}$ is said to be 
acceptable if the associated differential form
\begin{equation}
\bar{W}(\sigma_{i (r)} \times \sigma_{j, (p)}).
\end{equation}
is differentiable globally (i.e in each hypercubes). \\

i) In such case, the co-homology is not violated by the introduction of 
the product $\times$. 

ii) Given a one chain and a $p$-chain, the interior product of the 
$p$-chain by the one chain gives rise to an acceptable product.

\end{lemme}
The statement i) stresses the importance of differentiability and ii) was established in the proof of Lemma \ref{lmm} 
\subsection{Alternative definition of the discrete analogues of $\star$ 
and $\wedge$ in the product space}
Let us now follow the logic of the introduction of the product 
$\times$, which was required by exactness. The natural question is of its 
applicability to the other operations in the theory. In fact, we now find that 
if we use $\times$ to define a new Hodge star operator, it follows that 
the operation is exact, and there is no need for a dual complex and the Poincar\'e duality holds.  
\begin{defi} Hodge operation \\

The alternative definition of the discrete Hodge star reads
\begin{equation}
\label{stare}
\bigstar^E \sigma_{j, (p)} \doteq \sigma_{j , (p)} \times \sigma^{d(p)}
\end{equation}
where $d(p)$ is one of $(0, 1, 2, 12)$ referring to the generic constant forms. 
\end{defi}
This definition immediately leads to
\begin{lemme} Exactness of the Hodge star \\

The discrete Hodge star is exact
\begin{equation}
\bar{W} \bigstar^E = \star W.
\end{equation}
\end{lemme}
This statement can be proved by constructing the generic examples, as follows.
\begin{examp} Discrete Hodge star $\bigstar^E$ \\

The simplest example is provided by the constant forms:
\begin{equation}
\bigstar^E \sigma^0 = \sigma^0 \times \sigma^{12} = \sigma^{12},
\end{equation}
The identification is obvious since $\varphi^{(0)}W(\sigma^0)  = 
\varphi^{(0)} (1) = 1$. Thus,
\begin{equation}
\bar{W} ( \sigma^0 \times \sigma^{12} ) = 1 \wedge dx\wedge dy = 
\star W (\sigma^0)
\end{equation}
so the result is exact. Let us consider the more general case of one 
edge, i.e not a constant form. Now consider
\begin{equation}
\label{hodgy}
\bigstar^E [(i,j) (i+1,j)] = [(i,j)(i+1,j)] \times \sigma^2
\end{equation}
To evaluate this, we need to consider the two squares bounding the edge 
$[(i,j) (i+1,j)]$, the upper one is denoted $A$, and the lower one is 
denoted $B$. Since we use local coordinates $(x,y)$ regardless of the 
cell used, we introduce the unit function $1_A$ for each square. 
We wish to show that Eq. \ref{hodgy}, is exact, i.e that
\begin{equation}
\bar{W}( \bigstar^E [(i,j) ( i+1,j)]) = \star  W([(i,j) (i+1,j)]).
\end{equation} 
So,
{\small
\begin{equation}
\bar{W} (\bigstar^E [(i,j)(i+1,j)]) = \varphi^{(0)}((1-y) dx)1_A 
\wedge dy 1_K +\varphi^{(0)}(y dx) 1_B\wedge dy 1_K 
= (1-y) dy 1_A + y dy 1_B,
\end{equation}
\small}
which is correct in both squares $A$ and $B$ and is zero everywhere 
else.
\end{examp}
The proof of the lemma, consists in reiterating the example for zero 
chains and two chains.

\begin{lemme} $D^K$-Cohomology on $C_{(p)}(K; \, \mathbb{R})$ \\

The chain space is such that under the new Hodge star, 
\begin{eqnarray}
\bar{W}\bigstar^E C_{(2)}(K;\, \mathbb{R}) &<& B^{(0)} (K;\, \mathbb{R}), \\
\bar{W}\bigstar^E C_{(1)}(K;\, \mathbb{R}) &<& B^{(1)}(K;\, \mathbb{R}), \\
\bar{W}\bigstar^E C_{(0)}(K;\, \mathbb{R}) &<& C^{(2)}(K;\, \mathbb{R}) \cong 
B^{(2)}(K;\, \mathbb{R}).
\end{eqnarray} 
This for the hypercubic complex. 
\end{lemme}
Clearly this is a limitation if one is to use the chain space rather than the product space of chains.
Also, consider the property
\begin{equation}
\delta^K = (-1)^{n(p+1)+1}\bigstar^E d^K \bigstar^E.
\end{equation}
One finds easily that
\begin{equation}
\delta^K \sigma_{i, (2)} = (-1) \bigstar^E d^K \bigstar^E \sigma_{i, (2)} = 0
\end{equation}
for any $\sigma_{i, (2)}$. The reason is simple, the Whitney form 
$W(\sigma_{i, (2)})$ is constant in the region $\sigma_{i, (2)}$ and 
therefore since the Hodge star is exact, $d^K \bigstar^E \sigma_{(i, (2)} = 0$. We are then lead to represent fields themselves, in the product space, so not just 
two chains which are not sufficient as a basis for the co-homology with 
operation $\delta$ defined as above. In general we will consider fields 
such as 
\begin{equation}
\psi^K = \sum_{i,j} \psi_{ij} \sigma_{i, (p)} \times \beta_{j, (q)}. 
\end{equation}
Let us now return briefly to the interior product. Because of the 
exactness of the new Hodge star $\bigstar^E$ we can now derive the discrete version of the interior 
product using Eq \eqref{hiraf}, which was 
approximate in the original construction. To that end, we need to introduce a wedge product using the product $\times$. 

\begin{defi} Discrete wedge $\bigwedge$ \\

Given two chains $\sigma$ and $\rho$. A wedge product is introduced 
\begin{equation}
\label{bigwedge}
\sigma_{i, (p)} \bigwedge \rho_{j, (q)} = \sigma_{i, (r)} \otimes 
\rho_{j, (p)} \times \sigma_{(p+q)}.
\end{equation}
where $\sigma_{(p+q)}$ is the appropriate constant form of degree $(p+q)$.
\end{defi}

This discrete wedge is numerically exact, by construction, it satisfies 
\begin{equation}
\bar{W} (\sigma_{i, (r)} \bigwedge \rho_{j, (p)} ) = W(\sigma_{i, (r)} 
) \wedge W(\sigma_{j,(p)}).
\end{equation}
We may now turn to the alternative definition of the interior product 
Eq. \eqref{hiraf}, the approximate discrete 
wedge was used, together with a definition of vector fields in the dual 
complex. The formula is numerically approximate. In the present 
approach we see that we may follow \cite{hirani, desb} in using the formula 
Eq. \eqref{hiraf}, but with the exact operations now defined in the product space, 
namely,

\begin{lemme} Relation to other formula for the interior product. \\

The discrete interior product may be expressed as
\begin{equation}
i_{X^K} \sigma_{(p)} = (-1)^{p (n-p)}\bigstar^E X^K \bigwedge \bigstar^E \sigma_{(p)}. 
\end{equation}
It is exact.
\end{lemme}
The cohomology is correct by virtue of the exactness of $d^K$ applied 
after the exact operation of $\bigstar^E$.

Before turning to the simplicial formulation, we consider briefly the Clifford product as it plays a role in the 
discussion of spinors. The product is 
\begin{equation}
\omega \vee \rho = \sum_{p} (-1)^{F(p)} (i_{\partial_{(\mu_1)}} \ldots  
i_{\partial_{\mu_p}} \omega ) \wedge i_{\partial_{\mu_1}} \ldots 
i_{\partial_{\mu_p}} \rho,
\end{equation}
this mapping involves both $i_v$ and the wedge product. We have defined 
both maps. Recall that the discrete wedge $\wedge^K$ is approximate so will be the 
Clifford product \cite{us}. We however found some exactness, the discrete version of the Clifford product
\begin{equation}
\sigma_{i, (p)} \vee \sigma_{j , (q)} \doteq \sum_{p} (-1)^p 
(i_{\sigma^{(\mu_1)}}^K \ldots  i_{\sigma^{(\mu_p)}}^K \sigma_{i, (p)} ) \wedge^K 
i_{\sigma^{(\mu_1)}}^K \ldots i_{\sigma^{(\mu_p)}}^K \sigma_{j, (q)}
\end{equation}
is exact if at least one of $\sigma_{i, (p)}$ and $\sigma_{j , (q)}$ is 
a constant form, i.e is one of $\sigma^0, \, 
\sigma^1, \, \sigma^2, \, \sigma^{12}$. In that case it satisfies
\begin{equation}
\sigma^{d} \vee \sigma_{j, (q)} = A ( W(\sigma^d) \vee W(\sigma_{j, 
(q)} ))
\end{equation}
and is exact.
\subsection{Simplicial formulation}
We now consider the simplicial case. It enables one to consider 
manifolds such as the sphere, but also sheds some light on the method. Now, 
given one such triangle $\sigma_{i, (2)}$, we introduce the simplicial 
version of the Whitney forms in the standard simplex coordinates:
\begin{equation}
W([0]) = 1-x-y, \;
W([1]) = x, \;
W([2]) = y.
\end{equation}
A subtle point is that these coordinates do not match the local 
coordinates of the embedding $\mathbb{R}^N$ up to an isometry. Since 
differential forms are coordinate independent objects this assignment is fine. 
Moreover, the contracted chain term orthogonal to the vector may be 
defined in terms of the orthogonality in the embedding $\mathbb{R}^N$ in 
order to keep geometrical data intact.

The one forms are:
\begin{equation}
W([01]) = (1-y)dx + x dy, \; W([02]) = (1-x)dy + y dx, \; W([12]) = 
xdy-ydx.
\end{equation}
The only two-form is
\begin{equation}
W([012])= dx\wedge dy.
\end{equation}
To construct a valid interior product for the simplicial complex, we 
need to introduce some notation, as there is no direct correspondence 
between form components (i.e a $dx$ or $dy$ term) and co-chain in contrast 
to the hyper-cubic setting. 
The first step is to note that the co-chain $C^{(p)}(K;\, \mathbb{R})$ contains the 
globally constant forms, these are easily constructed, by inspection,
\begin{eqnarray}
1 &=& W ([0] + [1] + [2]),  \\
dx &=& W([01]-  [12]), \\
dy &=& W([02] + [12]), \\
dx\wedge dy &=& W([012]).
\end{eqnarray}
We write the chains on the right-hand side of the equation above as 
follows
\begin{defi} Simplicial notation \\

The linear combination of chains which maps to the constant form 
\begin{equation}
dx^{J} \doteq dx^{j_1} \wedge \ldots \wedge dx^{j_r}
\end{equation}
under the Whitney map, is denoted $\sigma^{J}_{(p)}$.
The index $J$ stands for an ordered list of $\mathbb{R}^N$ coordinate 
indices. For example $J = \{ 1; \, 2\}$ gives $W (\sigma^{J}_{(2)}) = 
dx^1 \wedge dx^2 = dx\wedge dy$.
\end{defi}
A co-chain admits various presentations in the product space of chains as for the hypercubic version,
\begin{equation}
W (\sigma_{(p)} ) = \bar{W} ( \sigma'_J \times \sigma^{J}_{(p)}).
\end{equation}
the decomposition on the right hand side has the advantage of 
separating the various form components which appear in the simplicial case. 

\begin{lemme} Simplicial interior product \\

 A chain $\sigma_{(p)}$ is decomposed as follows
\begin{equation}
\sigma_{(p)} = \sigma_{J} \times \sigma^{J}_{(p)}.
\end{equation}
The simplicial operation, where $X$ is a one-chain, is 
\begin{equation}
i_X \sigma_{(p)} \doteq X_{\mu} \otimes \sigma_{\hat{\mu}J} \times 
\sigma^{J}_{(p-1)}
\end{equation}
where $\sigma_{\hat{\mu} J}$ is the chain giving the correct 
coefficient for the $W (\sigma^{J}_{(p-1)})$ form component so that the discrete 
interior product $i_X$, is exact: 
\begin{equation}
\bar{W} i_X  = i_{W(X)} \bar{W}. 
\end{equation}
and it is compatible with the other operations. 
\end{lemme}

In the formula for the discrete interior product, $X_{\mu}$ is the 
$\mu$-component of the chain corresponding to $X$, as we have seen, this 
component can be extracted. Next, the chain $\sigma_{\hat{\mu} J}$ appears in the sum $\sigma_{\hat{\mu} J} \times \sigma^{J}_{(p-1)}$ 
where $J$ is the summed index. The chain $\sigma_{\hat{\mu} J}$ is such 
that 
\begin{equation}
i_{\frac{\partial}{\partial x^{\mu}}} W( \sigma_{\hat{\mu}J} \times 
\sigma^{J}_{(p)}) = \bar{W} (\sigma_{\hat{\mu}J} \times \sigma^{J}_{(p-1)}). 
\end{equation}
To establish the validity of the formula, we note that
\begin{lemme}
The space of extended Whitney forms is invariant under contraction by a 
globally constant vector field.
\end{lemme}

To show the lemma, we may take the more general case of any dimensions, 
say $N$. Then the general formula for the Whitney form is usually 
written using $\mu_i \doteq W([i])$ as a notation, 
\begin{equation}
\label{eqlab}
W([i_1 \ldots i_r]) = \frac{1}{r!} \sum_k (-1)^k \mu_i d\mu_{i_1} \wedge 
\ldots \hat{d\mu_{i_k}} \wedge \ldots \wedge d\mu_{i_r}.
\end{equation}
Taking the following expression for the Whitney forms, with $x_i$ 
coordinates in the simplex, 
\begin{eqnarray}
\label{eqlab2}
\mu_i &=& x_i, \; \forall i \neq 0, \\
\mu_0 &=& 1 - \sum_i x_i.
\end{eqnarray}
With this list, it is found by inspection that the coefficient function 
of a form is expressible within the given simplex as a linear 
combination of Whitney zero forms, and the form components in Eq. \ref{eqlab} are a 
linear combination of constant forms.

\section{Further prescription for metric}
\label{further}
\stepcounter{within}
\stepcounter{within2}
\stepcounter{within3}
\stepcounter{within4}
\subsection{The picture}
Consider the manifold constructed 
using the standard simplex as the local chart, we find that this manifold 
$\mathfrak{S}^E$ has only a local flat metric $\delta_{ij}$ 
corresponding to normal coordinates in curved geometry, but if the surface has 
curvature, the metric is not globally defined in a smooth way, it changes 
discontinuously from simplex to simplex, which are global objects. In the Regge approach, 
the vielbein $\{ e^{(a)},\, a= 1,2\}$ is associated to edges and the change 
in the metric 
\begin{equation}
g = \delta_{a, b} e^{(a)} \otimes e^{(b)}
\end{equation}
from a simplex $\sigma_{i, (2)}$ to a neighbouring simplex $\sigma_{i+1, 
(2)}$ is defined by a matrix, which gives a finite rotation of the 
vector,
\begin{equation}
e^{(a)}(\sigma_{i, (2)}) = \Lambda^a_b e^{(b)} (\sigma_{i+1, (2)}),
\end{equation}
and hence the metric does not vary smoothly, i.e $dg$ is not defined 
globally, or we would say here, that it cannot be represented smoothly as 
a Whitney form.

We propose a solution by defining the vielbein smoothly per top dimensionality simplex, 
which means in turns that it is smoothly varying, i.e that within a 
triangle the metric is not locally flat as in the Regge 
approach. To that end, the product space introduced as a mean to represent the 
discrete interior product will prove crucial.

Let us start with the manifold structure we are given by the standard 
simplex construction. Given the complex $K$, we have a piecewise linear 
manifold with local chart given by the standard simplex. 
The simplicial complex is a geometrical approximation to the manifold $S$. The discrete metric in the product space of chains could be required to approximate this induced metric (it need not be the case), and capture local holonomies and other topological 
features as we will see.

For consistency, integrals will be performed in this chart, so for 
example:
\begin{equation}
vol_E (\sigma) \doteq \int_{|\sigma|} \star^E 1 = \int_{| \sigma |} dx 
\wedge dy.
\end{equation}
Of course, this Euclidean volume is not the volume in the new metric 
which is constructed algebraically below, the volume in the new metric is
\begin{equation}
vol_\theta (\sigma) \doteq \int_{|\sigma|} \star^\theta 1 = \int_{| 
\sigma|} \bigwedge_{i} \theta^{(i)}.
\end{equation}
We will favour the standard simplex coordinates, and will express 
\begin{equation}
\theta^{(a)} \doteq \theta_x (x,y) dx + \theta_y (x,y) dy,
\end{equation} 
in terms of Whitney forms. Also, we will use the new Hodge operation 
$\bigstar^E$ for $\star^E$ and at the end of this section we will define the discrete analogue of $\star^{\theta}$ as $\bigstar^{\theta}$.

Since the discrete version of the operations $\{\wedge, d ,\star, 
i_v\}$ is well-defined and compatible with the de Rham co-homology, 
there is a strong suggestion to use these to build the vielbein metric. 
Ounce an ansatz is proposed for the metric and the Cartan structure 
equations are written, the rest of the construction follows, since the 
covariant derivative $d_{\omega}$ is then fixed, and so is the curvature. 
In the standard simplex coordinates, an equation normalizes the Euler 
class and hence determines partly the metric.
\subsection{Vielbein}
To construct a generic metric, we start with the flat metric 
\begin{equation}
ds^2 = dx^2 + dy^2,
\end{equation}
it may be written immediately in the non-coordinate basis $\theta^{(1)} 
= dx, \theta^{(2)} = dy$. There are two different considerations that 
can be made, namely a change of either {\bf i)\bf} metric; or a change 
of {\bf ii)\bf} frame, i.e a change of coordinates. {\bf ii)\bf} 
corresponds to a gauge transformation.

Consider {\bf i)\bf}, since we have chosen a preferred coordinate 
system, namely the standard simplex coordinates, which is 
fixed, it is natural to consider the new metric $g = \theta^{(a)} 
\otimes \theta^{(a)}$, where $\theta^{(a)} = P dx^{a}$, while keeping the 
coordinate system fixed ( {\bf ii)\bf} is the alternative). It is of 
crucial importance that the constant forms $dx^{a}$ may be represented in 
the simplicial and hypercubic theories, i.e $A(dx^a) \in C_{(1)}(K)$ it 
is not an approximation. Also note that one is deforming a metric locally, but as we have seen, the topological features are global and require a normalization.

For {\bf ii)\bf}, the change of coordinates $x' = Px$, leads to the new 
frame $e' = e P^{-1}$, so that $\theta' = P \theta$, the connection 
transforms as 
\begin{equation}
\omega' = P' \omega P'^{-1} + P' d P'^{-1}, 
\end{equation}
while the curvature two form transforms as 
\begin{equation}
\mathcal{R}' = P' \mathcal{R} P'^{-1}.
\end{equation}

\subsubsection*{A piecewise flat vielbein}
We take the standard simplex coordinates $(x,y)$ as the preferred ones 
to express the vielbein. Let us consider the simplest implementation of 
metric. In the standard simplex coordinate system leads we have the Cartesian frame 
\begin{equation}
\bf{e} = \{ \partial_x, \, \partial_y \}.
\end{equation}
It is then natural to consider a new frame $e'$ defined via a change 
of coordinates to $x'= P x$, where $P$ is taken to be piecewise 
constant, the new frame is related to $e$ by  
\begin{equation}
e' =  e  \frac{\partial x}{\partial x'} = e P^{-1},
\end{equation}
however since the dual basis $\theta$ satisfies 
\begin{equation}
<\theta^{(a)}, e_{(b)} > = \delta^{a}_{b}.
\end{equation}
it is immediate that $\theta'= P\theta  $. First let us introduce the 
new coordinates
\begin{eqnarray}
\alpha^{(1)} (1-x - y) + \beta^{(1)} x + \gamma^{(1)} y &=& x' \\
\alpha^{(2)} ( 1 - x - y ) + \beta^{(2)} x + \gamma^{(2)} y &=& y'.
\end{eqnarray}
To derive the new frame, in $(x',\, y')$ coordinates, we derive $P$ 
form the above equation and fix $P \theta$ as the vielbein in the old 
coordinates $(x,y)$, so that it is a new metric (corresponding to {\bf i) \bf} 
above).

Since $P$ is constant in the interior of the standard simplex 
$\phi_{\triangle} \sigma$, it follows from the first Cartan structure equation 
that 
\begin{equation}
d P \theta = - \omega \wedge P\theta = 0,  
\end{equation} 
so the connection $\omega$ is piecewise constant. Hence the metric is 
flat in the interior of the cell. We have just proved the following

\begin{lemme}{Family of flat metrics} \\

The space of chains gives rise to a family of metrics which are flat 
in the interior of each simplex. 
\end{lemme}
Consider $P$ to be a constant non-singular matrix piecewise defined. Of course this implies that the new frame $e'$ gives a 
piecewise flat metric, with piecewise constant connection, and vanishing 
curvature two-form.
\subsubsection*{Generic metric}
In order to construct a non-zero valued curvature two-form, we need to 
construct a non-piecewise constant $P$. To do that in a consistent way, 
we will use the discrete operations introduced above. 

The generic $2D$ metric is 
\begin{equation}
ds^2 = dx^2 + G^2(x,y) dy^2.
\end{equation}
The vielbein is $\theta^{(1)} = dx,\, \theta^{(2)} =  G(x,y) dy$. and 
so $P^1 = 1, \, P^2 = G(x,y)$. Immediately one obtains the connection 
and curvature 
\begin{eqnarray}
\omega_{12} &=& - \partial_{x} G(x,y) dy, \\
\mathcal{R} &=& (\partial_x)^2 G(x,y) dx \wedge dy.
\end{eqnarray}

The parameters $(x,y)$ are the standard simplex coordinates. Note that 
the natural choice of the vielbein as a one-chain will not do because 
two derivatives give zero (because it is a Whitney one-form), as in the case of constant $P$. To move 
forward, we take the following

\begin{defi}{Ansatz for the vielbein} \\

Let the discrete vielbein be
\begin{eqnarray}
\label{ansatz}
\theta^{(1)} &\doteq& A(dx) \\
\theta^{(2)} &\doteq& i_{X} \bigstar^E i_{Y} A(dy). 
\end{eqnarray}
The two vectors $X$ and $Y$ are arbitrary one chains, up to giving Euclidean 
signature, and parametrize a family of metrics with line element
\begin{equation}
ds^2 = A(dx)\otimes A (dx) + i_{X} \bigstar^E i_{Y} A(dy)\otimes i_{X} 
\bigstar^E i_{Y} A(dy).
\end{equation}
\end{defi}

To obtain an orthogonal metric we require that $X$ and $Y$ be such that 
\begin{equation}
\label{constraint}
i_X \bigstar^E i_Y dy = G^K \times  A(dy),
\end{equation}
for some chain $G^K$ to be determined, since $dx$ and $dy$ are 
orthogonal in the standard simplex coordinates. The ansatz for the vielbien 
should be understood in terms of connection and curvature.
\begin{lemme}{Discretized Cartan structure equations} \\

The discretized version of the torsionless Cartan structure equations is given by
\begin{eqnarray}
d^K \theta^{(a)} + \bigstar^E i_{\omega^K}\bigstar^E \theta^{(b)} &=& 
0, \\
\label{covar}
d^K \omega^K + \bigstar^E i_{\omega^K} \bigstar^E \omega^K &\doteq& 
\mathcal{R}^K.
\end{eqnarray}

The ansatz allows to solve the first equation, the second is a 
definition of the curvature two-form, the metric is normalized by the 
Gauss-Bonnet theorem.
\end{lemme} 

To prove the lemma we will consider a generic example below, first we solve the problem in the continuum theory by calculating $\omega$. 
Now, choose $X$ and $Y$ such that Eq. \eqref{constraint} holds. Next,
\begin{equation}
d \theta^{(2)} =( \partial_x X_x  Y_y +  X_x \partial_x Y_y) dx\wedge 
dy,
\end{equation}
since the second derivative $(\partial_x)^2$ gives zero on chains $X$ 
and $Y$, we are left with
\begin{equation}
\omega_{21} = - ( \partial_x X_x  Y_y +  X_x \partial_x Y_y) dx.
\end{equation}
Next, in 2D the curvature two-form is simply
\begin{equation}
\mathcal{R} = d \omega = 2 \partial_x X_x \partial_x Y_y dx \wedge dy.
\end{equation}
Finally, the volume form is
\begin{equation}
\star^\theta 1 = \theta^{(1)} \wedge \theta^{(2)}.
\end{equation}
The curvature two-form $\mathcal{R}$ is piecewise 
constant, it is therefore a two-chain and does not require the product space to be represented. 

Up to now we have defined the discrete objects, and solved the continuum version of the first Cartan structure equation for the connection. We now introduce the covariant derivative.

\begin{defi}{Covariant derivative and curvature} \\

Define the covariant derivative $d_{\omega}$, as 
\begin{equation}
d_{\omega^K} \doteq d^K + \bigstar^E i_{\omega} \bigstar^E,
\end{equation}
Then the curvature two-form is given by
\begin{equation}
\mathcal{R}^K (\omega^K) \doteq (d_{\omega^K})^2 
\end{equation}
mimicing the continuum formula $\mathcal{R} (\omega) = d_{\omega}^2$ and Eq. \eqref{covar}.
\end{defi} 

We note immediately that 
\begin{lemme}{Exactness of $d_{\omega}$ and $\mathcal{R}$} \\

The discrete covariant derivative is exact and can be expressed as 
\begin{equation}
\bar{W}d_{\omega^K} = d_{W(\omega^K)}.
\end{equation}

In turn, the curvature is also exact:  
\begin{equation}
\mathcal{R} ( \bar{W}(\omega^K)) = \bar{W} (\mathcal{R}^K (\omega^K)).
\end{equation}
\end{lemme}
The lemma implies that the curvature two form is the true curvature 
corresponding to the connection $\omega^K$. 

To consider parallel transport, 
\begin{equation}
d_{\omega} \rho = 0,
\end{equation}
one may write equations similar to the first Cartan equation.

To determine $X, \, Y$ in the ansatz for the vielbein, we find that 
these are constrained by topology, we must impose the Gauss-Bonnet 
theorem, depending on the number of independent variables $Y$, this is a 
constraint that does not determine the volume of the surface in the new 
metric. 

To make contact with topology, we must calculate the first Pontryagin 
class $p_1(K)$ and show that the Euler class gives the correct Euler 
characteristic upon integration. Now,
\begin{equation}
p_1 (K) = \frac{1}{8 \pi^2} tr \mathcal{R}^2.
\end{equation}
but we have computed explicitly $P$, so the curvature is modified to 
$P^{-1} \mathcal{R} P$, while if we consider the spherical metric $ds^2 = 
(d\theta)^2 + (\sin \theta)^2 (d \phi)^2$, we have a new $P= P'$ such 
that $\mathcal{R} = P'^{-1} \mathcal{R}_{sphere} P'$, and by cyclicity 
of the trace, we get the same result, but only locally, i.e as we discussed, this argument does not hold globally and so we appeal to a normalization for the Euler class. We return to this point in the next section where an explicit spherical topology example is worked out.

Let us now turn to a simple example, to illustrate the method in more 
detail.

Having introduced a map $\bigstar^E$ which corresponds to the Euclidean
Hodge star, we are now lead to consider the discrete Hodge star for the
generic metric $\bigstar^{\theta}$. 
   After application of the curved metric Hodge operator one obtains a multiplicative $\sqrt{g}$ factor, this factor is a determinant and determines an orientation like the volume element on a Riemannian manifold 
\begin{equation}
vol = \sqrt{g} dx^{1} \wedge \ldots \wedge dx^{N},
\end{equation}
which has "outer orientation".

We have proposed a new discrete Hodge star which extends the former one for a curved metric which does not require the introduction of the dual lattice and handles the Hodge star in a way that has the correct algebraic properties. We now proceed to describe it in details. We start with the volume element
\begin{equation}
vol = \theta^{(0)} \wedge \ldots \wedge \theta^{(3)}.
\end{equation}
After discretisation, it may be written as
\begin{equation}
vol^K = \bar{W} ( G^K \times \sigma^{012}).
\end{equation}
In that case the factor $G^K$ is the discrete analogue of $\sqrt{g}$. The discrete Hodge star is then given by a map from the product space of chains onto itself. 
\begin{defi} Discrete Hodge star $\bigstar^{\theta}$
\begin{equation}
\label{hod}
\bigstar^{\theta} = G^K \times \bigstar^E.
\end{equation}
And 
\begin{equation}
(\bigstar^{\theta})^{-1} = (G^K)^{-1} \times \bigstar^E. 
\end{equation}
The inverse $(G^K)^{-1}$ is only formally defined \footnote{It cannot be represented exactly as a chain or product chain, due to the functional form of Whitney elements.} in the following sense:
\begin{equation}
W((G^K)^{-1}) \doteq (W (G^K))^{-1}.
\end{equation}
\end{defi}
One verifies that indeed
\begin{eqnarray}
\bar{W}(\bigstar^{\theta}(\bigstar^{\theta})^{-1}) &=& \bar{W}((\bigstar^{\theta}) (\bigstar^{\theta})^{-1}) \\
&=& W(G^K) W((G^K)^{-1}) \\
&=& 1.
\end{eqnarray}
We also verify that the new Hodge star is related to the vielbein basis,
\begin{eqnarray}
\bigstar^{\theta} 1 &=& \theta^{(1)} \wedge \theta^{(2)},\, 
\bigstar^{\theta}  \theta^{(1)}\wedge \theta^{(2)} = 1 \\
\bigstar^{\theta} \theta^{(1)} &=& \theta^{(2)}, \, \bigstar^{\theta} 
\theta^{(2)} = - \theta^{(1)}.
\end{eqnarray}

Now, the Hodge star is primarily used to define the coboundary operator:
\begin{equation}
\delta  \doteq \star^{-1} d \star.
\end{equation}
Which in the present formalism should be discretised as
\begin{equation}
\delta^{\theta} = (G^K)^{-1} \times \bigstar^E d  G^K \times \bigstar^E. 
\end{equation}
In this context, the discrete exterior derivative is piecewise exact, while the co-boundary operator is piecewise approximate for curved metric, since for the flat case, the Euclidean Hodge star $\bigstar^E$ is exact and in that case $(\bigstar^E)^{-1} = \bigstar^E$.

To establish the identity $(\delta^{\theta})^2 =0$, we calculate
\begin{eqnarray}
(\delta^{\theta})^2 &=&    (G^K)^{-1} \times \bigstar^E d  G^K \times \bigstar^E (G^K)^{-1} \times \bigstar^E d  G^K \times \bigstar^E \\
&=& 0 
\end{eqnarray}  

 Next, let us turn to the curved metric Laplacian. For a general metric, in the continuum, on zero forms for example
\begin{equation}
\triangle f = -\frac{1}{\sqrt{g}} \partial_{\nu} \sqrt{g}g^{\nu \mu} \partial_{\nu}f.
\end{equation} 
It is specified in the ring of operations we have introduced as 
\begin{equation}
\triangle^K = D^K \delta^K + \delta^K D^K,
\end{equation}
where 
\begin{equation}
\delta^K = \bigstar^{\theta} D^K \bigstar^{\theta}.
\end{equation}  

\subsection{Explicit construction for a complex with spherical topology}


\subsubsection*{Choosing a discrete metric parametrization $X$ and $Y$}
For simplicity we take the boundary of the tetrahedron as a triangulation of $S^2$. The faces are $[012]$, $[213]$, $[320]$, $[013]$. We make an ansatz for the vectors $X, Y$:
\begin{eqnarray}
Y &=& \sum_{i < j} Y_{ij} [ij], \\
X &=& i_{\partial_y} Y \times A(dy).
\end{eqnarray}
ans use the general formula given above for the discrete vielbein. We also introduce a normalization for the vielbein $\theta^{(a)} \doteq N \theta^{(a)}$. This normalization does not appear in the formula for the spin connection $\omega$ as given in the Cartan structure equation. It will allow us to fix the overall volume of the sphere.

Next, using the explicit form of the Whitney one-forms, we find
\begin{eqnarray}
\label{not}
Y_y &=& Y_{01} x + Y_{02} (1-x) + Y_{12} x, \\
Y_x &=& Y_{01} (1-y) + Y_{02} y - Y_{12} y.
\end{eqnarray}
By direct computation, this leads to 
\begin{eqnarray}
\theta^{(2)} &=& N(Y_{01} x + Y_{02} (1-x) + Y_{12} x)^2 dy \\
\omega_{21} &=& - 2 (Y_{01} x + Y_{02} (1-x) + Y_{12}x  )(Y_{02} - Y_{01} - Y_{12})dy \\
\mathcal{R} &=& -2 (-Y_{01}  + Y_{02} - Y_{12})^2 dx \wedge dy. 
\end{eqnarray}
We are computing as in the continuum theory, but with Whitney forms, because the formula for the discrete Cartan structure equation and exactness of $d^K$ allows such identification.

\subsubsection*{Killing vectors} 
The equation is 
\begin{equation}
L_Z g = 0.
\end{equation}
That is
\begin{equation}
L_Z \theta^{(1)} \otimes \theta^{(1)} + \theta^{(1)} \otimes L_Z \theta^{(1)} + L_Z \theta^{(2)} \otimes \theta^{(2)} + \theta^{(2)} \otimes L_Z \theta^{(2)} = 0 .
\end{equation}
So let $Z = \sum_{ij} Z_{ij} [ij]$, and let us look at the triangle $[012]$, then
\begin{eqnarray}
\label{kun}
L_Z \theta^{(1)} &=& (-Z_{01} + Z_{02} -Z_{12})dy,  \\
\label{kdeux}
L_Z \theta^{(2)} &=& 2 Y_y Z_x (Y_{01} - Y_{02} + Y_{12}) dy + (Z_{01} - Z_{02} + Z_{12}) Y_y^2dx
\end{eqnarray}
Where we follow the notation $Z_x$, $Z_y$ as in Eq. \eqref{not}. The Killing equation are then
\begin{eqnarray}
-Z_{01} + Z_{02} - Z_{12} &=& 2 Y_y Z_x (Y_{01} - Y_{02} + Y_{12}) \\
(Z_{01} - Z_{02} + Z_{12}) Y_y^2 &=& 0 
\end{eqnarray} 
These equation cannot be solved exactly as they stand apart from trivial case. The de Rham $A$ can be applied to the components and then solved, otherwise we may minimize the expression $\bar{W} L_Z g$.  Also note that there are such equations as Eq. \eqref{kun}, \eqref{kdeux} for each simplex of the triangulation.
\subsubsection*{Euler characteristic}
To address the topological invariant we consider the simplest case we let $Y_{01} = Y_{02} = Y_{12} = Y$. Then,
\begin{equation}
\theta^{(2)} = (Y + Yx)^2dy,\;
\omega_{12} = 2( Y + Y x)Y dy,\;
\mathcal{R} = 2 Y^2 dx\wedge dy.
\end{equation}
Then, 
\begin{equation}
p_1 = \frac{1}{8 \pi^2} tr \mathcal{R}^2 
= \frac{1}{2\pi^2} (Y^2 dx\wedge dy)^2
\end{equation}
so the Euler class is the "square root" of this last expression,
\begin{equation}
e(M) = \frac{1}{\pi\sqrt{2}} Y^2 dx\wedge dy,
\end{equation}
it is a volume element itself, and satisfies the Gauss-Bonnet theorem
\begin{equation}
\int_{|M|} e (M) = \chi( S^2) = 2.
\end{equation} 
This determines $Y$, by integration, where $N_T$ is the number of triangles. So we have the following equation from topology,
\begin{equation}
\sum_i \int_{|\sigma_{i, (2)}|} e(M) =  N_T Y^2\frac{1}{\pi \sqrt{2}} =2. 
\end{equation}
where the area is calculated in the standard simplex coordinates as 
\begin{equation}
\int_{0}^{1} dx \int_{0}^{1-x} dy = \frac{1}{2}.
\end{equation}
So we found that
\begin{equation}
Y = \frac{\pi \sqrt{8}}{N_T}.
\end{equation}
We may now evaluate the volume of the surface $S$ with the metric $\{ \theta^{(a)} \}$,
\begin{equation}
vol_{\theta} (K) = \sum_i \int_{|\sigma_{i, (2)}|} \theta^{(1)} \wedge \theta^{(2)} = \frac{8 \pi^2}{N_T^2} \sum_i \int_{\sigma_{i, (2)}}   (1+x)^2 dx \wedge dy 
= \frac{28 \pi^2}{3 N_T}.
\end{equation} 
To limit the discussion to our example of one tetrahedron, $N_T = 4$ and we find 
\begin{equation}
vol_{\theta} (K) = (\frac{\pi}{N_T})^2.
\end{equation}
Also local holonomies may be considered, by taking a given triangle 
$[012]$. In its interior we write, what is simply the Stokes theorem (this 
is the 2D case, in  general the exponential is used to write the 
holonomy)
\begin{equation}
\int_{| [012] |} \mathcal{R} = \int_{| \partial [012] |} \omega.
\end{equation}
This concludes the discussion of metric.
\section{Relation to the non-commutative geometry}
\label{relation}

\stepcounter{within}
\stepcounter{within2}
\stepcounter{within3}
\stepcounter{within4}

As established by Woronowicz \cite{wor}, a non-commutative differential geometry is associated to a 
graph (see also \cite{Dimakis:1994qq}), it is a deformed calculus with deformation parameter of 
the order of the lattice spacing. The typical graph that determines such calculus is a Hasse diagram, i.e a graph with arrows.

Consider the commutative group lattice $\mathcal{L}$ (the vertices of $\mathcal{L}$ coincide with the vertices of the hypercubic complex $K$), where the group operation is 
given by translation by one lattice spacing. A Hopf algebra can be 
constructed in such system (in the context of physics see \cite{Aschieri:2002pn}), and the Yang-Baxter 
equations are found in terms of the discrete wedge product, concretely, as we recall below, the non-commutativity is between differentials and functions. To start, Let us now compare the functions on the group lattice with the Whitney 
zero-forms.
\subsection{Functions}
 The basis of functions in the non-commutative formulation is $\{ e^x, \, x \in \mathcal{L} 
\}$, where $x$ labels a vertex of the group lattice $\mathcal{L}$. The algebra of 
functions is
\begin{eqnarray}
\label{e1}
e^x (y) &\doteq& \delta^{xy}, \\
\label{e2}
e^x e^y &\doteq& \delta^{xy} e^x.
\end{eqnarray}
A trivial property is $\sum_x e^x = 1$.
On the other hand, the Whitney zero forms, $\{\mu_i, \, i \in K\}$ are 
such that
\begin{equation}
\mu_i (j) = \delta_{ij}
\end{equation}
 which corresponds to Eq. \eqref{e1} and also satisfy $\sum_i \mu_i = 1$.
 
The second property Eq. \eqref{e2} may be written as \footnote{A vertex in 
the non-commutative picture is denoted $x$, in the present formalism it is denoted by $i$ as an index and by $[i]$ for a chain.}
\begin{equation}
[i] \wedge^K [j] = \delta_{ij}[j] 
\end{equation}
This gives a natural correspondence with the definition on a group lattice Eq. \eqref{e2}. We will see 
shortly that such an analogy is limited to zero forms. Next, we introduce a morphism $\mathbf{\rho}$, by first setting
\begin{eqnarray}
\label{rho}
\rho &:& \mathcal{L} \longrightarrow C_{(0)}(K, \mathbb{R}) \\
\rho &:& e^x \mapsto \mu_i.
\end{eqnarray}
where in this case $x$ and $[i]$ coincide. Now we define
\begin{equation}
\rho( e^x e^y) \doteq \delta^{xy} \rho(e^y).
\end{equation}
which make $\rho$ a mapping to the space of forms.
\subsection{Forms and the morphism $\rho$}
To develop further the correspondence we need to consider higher forms. 
At this point the non-commutativity between functions and forms appears in the non-commutative theory.
We specialise to the symmetric lattice \cite{Dimakis:1994qq} for reasons that will become clear below. 

The basis list of one forms consists of
\begin{eqnarray}
\theta^{\mu} &\doteq& \sum_x e^x d e^{x+\mu}, \\
\theta^{-\mu} &\doteq& \sum_x e^x d e^{x-\mu},
\end{eqnarray}
where $\mu$ is the unit vector in the $\mu$ direction on the group lattice.
This choice is special in the sense that it keeps locality, $e^x$ and 
$e^{x\pm \mu}$ are nearest neighbors. To define higher degree forms one needs to define the wedge product for this non-commutative basis,
\begin{equation}
\theta^{\mu}  \wedge \theta^{\nu} \doteq \theta^{\mu} \otimes \theta^{\nu} - \theta^{\nu + \mu} \otimes \theta^{\mu}.
\end{equation}
This operation leads to the braid matrix (see for example \cite{Aschieri:2002pn}). In the sequel we drop the symbol $\otimes$ for the tensor product of non-commutative forms.

The non-commutativity property of the product of function and one form is
\begin{equation}
e^{x \mp \mu} \theta^{\pm \mu} = \theta^{\pm \mu} e^{x}.
\end{equation}
This algebra of functions and forms is non-commutative so the 
correspondence with de Rham differential forms is apparently lost since the 
order in which the various terms are presented corresponds to different 
differential forms.
To remove this ambiguity, we note that there we can choose a preferred presentation of 
non-commutative forms that is unique
\begin{equation}
\omega = \sum_{i_1 < \ldots < i_r} e^{x_{i_0}} d e^{x_{i_1}} \ldots d e^{x_{i_r}}.
\end{equation}
We now make a prescription for $de^{x_{i_p}}$ by extending the 
range of $\rho$ from functions to forms. Since in the non-commutative setting the discrete exterior derivative denoted $d$ satisfies the Leibniz rule 
\begin{equation}
d (e^x e^y) = d e^x e^y + e^x de^y,
\end{equation}
we extend the morphism $\rho$ to accommodate this property,
\begin{equation}
\rho (d (e^x e^y)) \doteq d( \rho(e^x) \rho(e^y)) = d(\mu_i\mu_j) = 
d\mu_i \mu_j + \mu_i d \mu_j.
\end{equation}
The last equal sign is the continuum Leibniz rule on Whitney zero-forms. So we may 
now write the correspondence for any form, using the functorial property 
of $d$, we simply set
\begin{equation}
\label{canonical}
\rho ( \sum_i e^{x_{i_0}} d e^{x_{i_1}} \ldots d e^{x_{i_r}}) \doteq  
\sum_j \mu_{j_0} d \mu_{j_1} \wedge \ldots  \wedge d \mu_{j_r}.
\end{equation}
So we have found that under the map $\rho$, maintaining the Leibniz 
rule leads to the prescription
\begin{equation}
\rho (d e^x) = d \rho (e^x).
\end{equation}
Note that the form of Eq. \eqref{canonical} is reminiscent of the simplicial 
Whitney form expression, Eq. \eqref{eqlab} but it does not correspond to 
the hyper-cubic Whitney form. To establish a relation we introduce a new 
map $T$ in the following
\begin{lemme} The truncation map $T$ \\

For a hyper-cubic complex, the Whitney form may be expressed as:
\begin{equation}
\label{trunca}
W(\sigma_{i, (r)}) = T \sum_k (-1)^k \mu_i d\mu_{i_0} \wedge \ldots \wedge 
\hat{d\mu_{i_k}} \wedge \ldots \wedge d \mu_{i_r}
\end{equation}
where $T$ acts on the function coefficient, any factor in the 
expression Eq. \eqref{trunca} is of the form $(x^{\mu})^K$ or $(1-x^{\mu})^K$, where 
$K$ is a positive integer. The truncation map is such that each term of 
the form $x^{\mu}$ or $(1-x^{\mu})$ shows ounce, i.e any $K$ is set to 
unity.
\end{lemme}
To illustrate the map $T$, choose a hypercube, and maps its vertices to $\mu_i$ functions, for 
example $[01]$,
\begin{eqnarray}
T( \mu_0 d \mu_1 - \mu_1 d \mu_0) &=& T((1-x) (1-y) d((x (1-y) ) - x(1-y) d ((1-x)(1-y))) \\
&=& (1-y) dx,
\end{eqnarray}
which is indeed $W([01])$.
So the truncation map allows to recover a Whitney form. This allows to 
show the relation between the non-commutative and the commutative 
settings. The maps are

\begin{equation}
\label{diag}
\begin{array}{ccc}
 \Omega (\mathcal{L};\, \mathbb{R})              &
\stackrel{\rho}{\longrightarrow} &
\Omega (M;\,\mathbb{R})                          \\
           & \searrow{\phi}  &                 
\Big\downarrow{T}                           \\
 &  %
&
C(K;\, \mathbb{R}) 
\end{array}
\end{equation}
where 
\begin{equation}
\phi \doteq T \, o \, \rho,
\end{equation}
the point is that the composition of the two maps is well-defined and hence the relation to the non-commutative basis can be made.
The space $\Omega (\mathcal{L};\,\mathbb{R})$ is the non-commutative one. To 
construct the non-commutative space of forms, we start from 
the co-chain space and invert the maps. In 2D, zero forms in the square 
$[0123]$ are given by
\begin{eqnarray}
\mu_0 &=& (1-x) (1-y) = T \rho e^0, \\
\mu_1 &=& x(1-y) = T \rho e^1,      \\
\mu_2 &=& xy = T \rho e^2, \\
\mu_3 &=& (1-x) y = T \rho e^3.
\end{eqnarray}
Where $T$ plays no role as no truncation is needed. Next, for one-forms 
recall the example above.
\begin{eqnarray}
W([01]) &=& (1-y) dx = T (1-y)^2 dx = T (\rho e^0 d e^1 - e^1 d e^0), 
\\
W([12]) &=& x dy = T x^2 dy = T \rho (e^1 d e^2 - e^2 d e^1), \\
W([32]) &=& y dx = T y^2 dx = T \rho (e^3 d e^2 - e^2 d e^3), \\
W([03]) &=& (1-x) dy = T (1-x)^2 dy = T \rho (e^0 d e^3 - e^3 d e^0),
\end{eqnarray}
and the only two-form is
\begin{equation}
W([0123]) = \frac{1}{3}T(W([01]) \wedge W ([03]) + W ([32]) \wedge W([12]) + 
W([01]) \wedge W([32]))
\end{equation}
The important point here is that the truncation map projects a general 
form spanned by a subspace of $\Omega(\mathcal{L}; \, \mathbb{R})$ onto a subspace of forms 
which matches co-chains. The truncation map is invertible, i.e we can 
recover the form that is in one-to one relation with a non-commutative form 
(see Eq. \eqref{canonical}).

\subsection{Operations related to contraction}
The way we constructed the contraction was to involve a product space 
of chains. Such details are not needed when the contraction only 
involves constant vector fields, namely the dual $\frac{\partial}{\partial 
x^{\mu}}$ of $dx^{\mu}$ in the Euclidean embedding space.
We write the interior product as $i_{\sigma^{\mu}}$ which is identified 
with $i_{\frac{\partial}{\partial x^{\mu}}}$.
In that case, given a chain $\sigma_{(p)}$, we  find
\begin{equation}
i_{\sigma^{\mu}} \sigma_{(p)} = \eta_{\sigma^{\mu}} (\sigma_{(p)})
\end{equation}
which is a linear combination of $(p-1)$ chains.
The prescription for the relation to the non-commutative setting (see for example \cite{kanamori}) is the following
\begin{equation}
\label{contractmatch}
\eta_{\sigma^{\mu}} (\sigma_{(p)}) = \rho ( e_{\mu} 
\rho^{\star}(\sigma_{(p)})),
\end{equation}
where $\rho^{\star}$ is the pull-back map, it associates a unique 
non-commutative differential form to a chain. The map $e^{\mu}$ of the 
non-commutative theory is
\begin{equation}
e_{\mu} \doteq \frac{\partial}{\partial \theta^{\mu}}  
\stackrel{\rightarrow}{T}_{- \mu} + \frac{\partial}{\partial \theta^{-\mu}}  
\stackrel{\rightarrow}{T}_{\mu}
\end{equation}
where 
\begin{equation}
\stackrel{\rightarrow}{T}_{\pm \mu}f([(i,j)] = f([(i,j) \pm \hat{\mu})]).
\end{equation}
The displacement operator being inserted to compensate for the 
non-commutativity involved in swapping the function and form before operating 
the derivative $\frac{\partial}{\partial \theta}$. It has been shown 
that the contracted form in the non-commutative setting is free of 
displacement operators. This is a hint of the matching we expect in Eq.
\eqref{contractmatch}. \\
\noindent
We now move to the discussion of the Clifford product. Consider the square $[0123]$, and local coordinate chart $(x,y)$. We have four functions $e^0,\ldots,e^3$ and four one forms $de^0,\ldots, 
de^3$. Consider the form $e^0 \theta^x$. Then,
\begin{equation}
i_{\partial_x} (\mu_0 d \mu_1 - \mu_1 d \mu_0 ) = i_{\partial_x} 
(1-y) dx 
= (1-y)= \mu_0 + \mu_1.
\end{equation}
find that the functional form given by the right hand side of 
\eqref{canonical} is preserved. 

Consider two one forms 
$f_{\mu} dx^{\mu}$ and $g_{\nu} dx^{\nu}$, their Clifford product is
\begin{equation}
f_{\mu} dx^{\mu} \vee g_{\nu} dx^{\nu} = f_{\mu}g_{\nu} 
\delta_{\mu,\nu} + f_{\mu}g_{\nu} dx^{\mu} \wedge dx^{\nu}
\end{equation}
For example,
\begin{equation}
W([01]) \vee W([01]) = \frac{\partial}{\partial dx} W([01]) \wedge 
\frac{\partial}{\partial dx} W([01]) = (1-y)^2.
\end{equation}
This is problematic, so let us pull it back to the non-commutative form 
basis:
\begin{equation}
\rho^{\star} (1-y)(1-y) = (e^{0}+ e^{1})(e^{0} + e^{1}) = (e^{0} +e^{1})
\end{equation}
We have to check that this truncation of the function is consistent 
with the associativity property of the Clifford product. Let us check the associativity with an example:
\begin{examp} Associativity of the Clifford product
\begin{eqnarray}
\rho^{\star}T^{\star}([01] \vee^K [02]) &=& \rho^{\star} T^{\star}([01] \wedge^K [02]) \\
&=& (e^0d e^1 - e^1 d e^0)(e^0 d e^2 - e^2 d e^1) \\
&=& e^0 de^1 de^2. \\
\rho^{\star} T^{\star} ([02] \vee^K [03]) &=& e^0 d e^2 de^3.
\end{eqnarray}
Then,
\begin{eqnarray}
[01] \vee^K ([02] \vee^K [03]) &=& e^0 d e^1 d e^2 d e^3, \\
([01] \vee^K [02]) \vee^K [03]  &=& e^0 d e^1 d e^2 d e^3.
\end{eqnarray}
\end{examp}
In general, there are two cases for the multiplication of functions as follows, 
the first one is when we are multiplying two identical functions like 
$(e^0 + e^1)$, in this case a truncation is needed i.e $T(1-y)^2 =(1-y)$, the other instance is when there are different $(e^1 + e^2)$ multiplied by $(e^1 + e^3)$ then there is no truncation i.e
\begin{equation}
(e^1 + e^2)(e^1 + e^3)= e^1
\end{equation}
is the correct result after mapping both sides under $\rho$.
A special case is when we take the Clifford product of two edges which 
are parallel within the cell. For example $[01]$ and $[32]$. In the 
original picture,
\begin{equation}
T^{\star}([01]\vee^K [32]) = (1-y) y,
\end{equation}
\begin{equation}
\rho^{\star} ((1-y)y) = ((e^0 + e^1)(e^3 + e^2)) = 0.
\end{equation}
Thus this destroy associativity? \\
 
To settle this point we take the edge connecting the two parallel ones, 
$[03]$ and consider the associativity.
\begin{equation}
\rho^{\star}T^{\star}([01]\vee^K ([32] \vee^K [03])) = (e^0+e^1) (e^3) = 0
\end{equation}
which is equal to
\begin{equation}
\rho^{\star}T^{\star}(([01]\vee^K [32]) \vee^K [03]) = ((e^0+e^1)(e^3 +e^2)) 
(e^0+e^3)) = 0
\end{equation}
So the associativity is kept.
The examples show that the correspondence between the Clifford product 
and its non-commutative version is as follows:
\begin{lemme}
The non-commutative Clifford product $\vee$ is related to the discrete 
product $\vee^K$ by application of the truncation map $T$,
\begin{equation}
 \rho^{\star} T^{\star} (\sigma \vee^K \eta) = \rho^{\star}((T^{\star} \sigma 
)\vee(T^{\star} \eta))
\end{equation}
the right hand side is an associative product, therefore so is the left 
hand side.
\end{lemme}
What this means for the relation to non-commutativity is that the operations 
of contraction and Clifford product defined here admit a well defined 
projection to the non-commutative operations on a group lattice, however 
the correspondence involves a truncation of the function space which 
violates the algebraic properties of the operations. 

This partial failure of the correspondence for the latest operations is conceptually no surprise. There is no satisfactory definition of these operations in the non-commutative picture. This applies also to the Hodge star. That is using the non-commutative geometry as a substitute for the ordinary differential calculus does not work.

\section{Discussion}
In order to test the usefulness of the method in concrete applications, we propose three concrete applications:
\begin{itemize}
\item 
Topological field theory: calculation of the Reidemeister torsion in a combinatorial way.
\item
Simplicial gravity: comparing with 2D random surface models for gravity, for Nambu-Goto and for actions including matter fields.
\item 
Electromagnetic theory: discretising Maxwell's equations. 
\end{itemize}
The first application is concerned with the calculation of the Redeimeister torsion \cite{ray} by explicitly constructing a simplicial regularization of the Abelian Chern-Simons theory. In \cite{Us} the geometric discretisation scheme using the dual complex based discrete Hodge star $\star^K$, $\star^L$ recalled above was used. It was shown in this context that 
\begin{equation}
\label{tor}
T = \sqrt{ \frac{1}{N_0 N_3} \det(\partial_1 d_0) \det (\partial_2 d_1)^{-1} \det(\partial_3 d_2)}.
\end{equation}
where $N_0$ is the number of vertices and $N_3$ the number of three simplices of the simplicial triangulation of the lens space or of $S^3$ as discussed in the above paper. The overall factor $\sqrt{\frac{1}{N_0 N_3}}$ is then the metric dependent factor of the expression for the combinatorial torsion. This factor can be either derived from the original expression of Ray-Singer, or can be deduced numerically when showing the subdivision invariance of the torsion.

In the present formulation, we expect the following functional form for the torsion
\begin{equation}
\label{tor}
T' = \sqrt{ F \det(\delta_1^K D_0^K)|_{\Phi^{\star}C(K; \mathbb{R})} \det (\delta_2^K D_1^K)|_{\Phi^{\star} C(K; \mathbb{R})}^{-1} \det(\delta_3^K D_2^K) |_{\Phi^{\star}C(K; \mathbb{R})}}.
\end{equation}
We have replaced the discrete operators of the geometric discretisation by their new version on the product space of chains, and we have added a factor $F$ to be specified, also note that each determinant is restricted to a certain subspace of the product space of chains.  To relate the two quantities, we relate the operators. Let us start with the first determinant term $\det(\partial_1 d_0)$ of $T$ and the corresponding one for $T'$. 

The key is the relation between the chain space $C(K; \mathbb{R})$ and the product space of chains. The latter is not a chain space, so we need to related it to the chain space $C(K; \mathbb{R})$. First we consider the mapping 
\begin{equation}
\Phi \doteq A \bar{W}: C_{(0)}(K; \mathbb{R}) \times C_{(p)} (K; \mathbb{R}) \longrightarrow C_{(p)}(K; \mathbb{R}).
\end{equation}
By inspection, the space $\ker \Phi$ is not empty. An example is a product chain such as $[0] \times [12]$. Next we introduce a pull-back map $\Phi^{\star}$ which gives a linear combination of product chains, this pull-back should be carefully chosen so that the equations below indeed hold (an example follows). The space $\Phi^{\star} C(K; \mathbb{R})$ is the topological representative of the chain complex in the product space. It can be shown that for any chain $\sigma_{i, (p)}$,
\begin{equation}
\Phi^{\star}(\partial^K_{p+1} d^K_p \sigma_{i, (p)}) = \Lambda \delta^K_{p+1} D^K_p  \Phi^{\star}( \sigma_{i, (p)}).
\end{equation}
The matrix $\Lambda$ corresponds to fixed degree dependent numerical factors. It follows that
\begin{eqnarray}
\det (\partial^K_{p+1} d^K_p) &=& \det(\Phi \Lambda_{p+1} \delta^K_{p+1} D^K_p \Phi^{\star}) \\
&=& \det(\Lambda) \det ' (\delta_{p+1}^K D^K_{p}) 
\end{eqnarray}
where $\det'$ is the determinant in the $\Phi^{\star}C_{(p)}(K; \mathbb{R})$ basis denoted as $\det(.)|_{\Phi^{\star} C(K;\mathbb{R})}$ in Eq. \ref{tor}.
Repeated application of the argument suggests that $T = T'$ if
\begin{equation}
F = \frac{1}{N_0 N_3} \det(\Lambda_3) \det(\Lambda_2) \det(\Lambda_1).  
\end{equation} 
Since this argument is somewhat abstract, we proceed with a calculation that illustrates the above calculations. Consider a single tetrahedron as a discretisation of $S^3$. We have shown that the operations on the product space are exact. To evaluate $\delta d$, we take one element of the chain basis $[12]\in C_{(1)}(K ; \mathbb{R})$. In the product space we take $\lambda_1 [1] \times [12] $ and $\lambda_2 [2] \times [12] \in C_{(0)}(K; \mathbb{R}) \times C_{(1)}(K; \mathbb{R})$. First we carry out the calculation in $C_{(1)}(K; \mathbb{R})$, it reads,
\begin{eqnarray}
d^K [12] &=& [312] + [012], \\
\partial^K d^K [12] &=& [12] - [32] + [31] + [12] - [02] + [01]. 
\end{eqnarray}
Now let us evaluate the analogue in the product space as
\begin{eqnarray}
\bar{W} ([1] \times [12]) &=& x ( xdy - y dx) \\
 \delta \bar{W} ([1] \times [12]) &=& 3xdy\wedge dy \\
 d\delta \bar{W} ([1] \times [12]) &=& -3 dy = 3\bar{W}([02] - [12] - [23]).
\end{eqnarray}
Next,
\begin{eqnarray}
\bar{W} ([2] \times [12]) &=& y ( xdy - y dx) \\
 d\delta \bar{W} ([1] \times [12]) &=& 3 dx = 3\bar{W} ([01] - [12] - [13]).  
\end{eqnarray}
By comparison of the calculation in the chain space with that in the product space, we find that if $\lambda_1 = -\lambda_2$, and adding the two terms, the product chain
\begin{equation}
\sigma = \lambda ([1] - [2] ) \times [12]
\end{equation}
is such that 
\begin{equation}
A ( \delta d \bar{W} (\sigma)) = 3 \lambda \partial^K d^K [12]. 
\end{equation} 
This factor of three will appear as an entry of the matrix $\Lambda$. What we have seen is how the chain space is captured in the product space and how the new operators $\delta^K$ and $D^K$ allow to compute the torsion. This needs to be confirmed numerically, furthermore, we may compute the determinants above in the entire product space to find the overall geometrical factor.

The second application is a test of the metric aspects of the theory and of the ability to express actions with the method. The theory of 2D gravity (string world sheet trajectories) is well understood, and a combinatorial solution of the problem has been constructed (see e.g \cite{amb}). It therefore provides a good test for discretising gravity before going to higher dimensions.  To formulate a valid analogue of the path integral 
formalism
in this theory, we first define the ensemble of all paths as the set of 
all
equivalence classes (under diffeomorphism) of surfaces X: S\( ^{1} \) 
\( \bigotimes  \)
I \( \rightarrow  \) \( \mathbb{R} ^{d} \) with boundaries \( \gamma _{1} \), 
\( \gamma _{2} \)
and we choose the action of a surface equal to its area:
\begin{equation}
S_{NG}\left( X\right) =\int dA\left( X\right) =\int d^{2}\xi \sqrt{\det 
h}
\end{equation}
$dA(X)$ denotes the area element of the surface in \( \Re ^{d} \) and $h$ 
is the
induced metric given by $h_{ab}=\frac{\partial X^{\mu }}{\partial \xi ^{a}}\frac{\partial X^{\mu 
}}{\partial \xi ^{b}}$, where (\( \xi ^{1},\xi ^{2}) \) are local coordinates on the manifold 
S\( ^{1} \)\( \bigotimes  \)
I. Using the Regge calculus, the Euler characteristic can be readily computed. In our approach, the metric is not the induced metric by embedding, as will discuss below. The partition function is defined by:
\begin{equation}
G\left( \gamma _{1},\gamma _{2}\right) =\int _{\left[ X\right] }D\left[ 
X\right] \exp \left( -S_{NG\left( X\right) }\right) 
\end{equation}
Where the integral is over the space of equivalence classes of 
surfaces. Consider a piecewise linear surface with action with chemical potential $\mu$, where $N_t$ is the number of triangles, the action is
\begin{equation}
S=\mu N_{t}.
\end{equation}
The partition function is:
\begin{equation}
Z(\mu)=\sum _{T\in \Gamma }\frac{1}{C_{T}}\exp \left( -\mu N_{t}\left( 
T\right) \right) 
\end{equation}
where $C_T$ is a symmetry factor of the triangulation. We wish to calculate expectation value of a physical observable $\Pi$. To that end, the Monte Carlo algorithm creates a sequence of A statistically independent configurations T\( _{a} \) with a probability distribution:
\begin{equation}
P\left( T_{a}\right) =\frac{1}{Z\left( \mu \right) 
}\frac{1}{C_{T_{a}}}\exp \left( -\mu N_{t}\left( T_{a}\right) \right) \end{equation}
and evaluates the expectation as:
\begin{equation}
<\Pi >=\frac{1}{A}\sum ^{a=A}_{a=1}\Pi \left[ T_{a}\right]
\end{equation}
where A is large. We expect a strict equality as $A\rightarrow \infty$ 
 The configurations are generated by a Markov chain \( t \) which can be 
regarded as a random walk in configuration space with transition function  
\begin{equation}
t\left[ T^{a},T^{b}\right] =C\min \left\{ 1,\frac{P\left[ T^{b}\right] 
}{P\left[ T^{a}\right] }\right\} 
\end{equation}
The calculation of observables such as the Hausdorff dimension of critical surfaces in the theory, is done by a Monte-Carlo simulation, or importance sampling. For a given genus, an algorithm generates random surfaces in the ensemble of surfaces with same topology, as determined by a probability for an update which is a functional of the action $S$. The latter will in general be expressed in terms of the Regge calculus. 

In the present case one proceeds differently. The metric provided on the surface is not the induced metric from embedding. The 2D metric is given by the prescription given in the section covering the metric above. Instead of modifying the surface by update as in the former method we propose to update the metric assignment $\{ Y_{ij} \}$. In this case the NG action reads
\begin{equation}
S_{NG}' = \int \theta^{(1)} \wedge \theta^{(2)}.
\end{equation}
The discrete version of it can readily be evaluated. It is 
\begin{equation}
S_{NG} = \sum_i \int_{|\sigma_{i, (2)} |} \bar{W} ( \theta^{(1)K}) \wedge \bar{W} (\theta^{(2)K}).
\end{equation}
To proceed, in each simplex one must assign an origin together with $x$ and $y$ axis. These are naturally associated to vertices in the standard simplex construction at work here. So for the two simplex $\sigma_{i, (2)}$, We relabel its three vertices as $O(i)$, $x(i)$ and $y(i)$ respectively.

The discretized action is
\begin{equation}
S = \sum_i \frac{1}{4} Y_{O(i)y(i)}^2 + \frac{1}{12} Y_{O(i) x(i)}^2 + \frac{1}{6} (Y_{O(i)x(i)} Y_{x(i) y(i)} + Y_{O(i) y(i)} Y_{O(i) x(i)} + Y_{x(i) y(i)} )
\end{equation}
The update in the ensemble of surfaces is now replace by an update in the ensemble of metrics parametrized by $\delta Y_{ij}= \pm 1$ for a fixed underlying surface. We will denote the updated metric as $Y \mapsto Y( \delta Y)$. After randomly choosing the edge $[ij]$, the probability for the update is a function of the corresponding change in the action. 

At each update, one must make sure that the Euler characteristic still holds. This can be done by updating the edges which have one vertex in common with either $[i]$ and $[j]$ in such a way that the topological identity still holds. To be precise, when the metric configuration is initialized, it is normalized so that the Euler characteristic is verified by integration of the Euler class. For the generic 2D metric the Euler class is 
\begin{equation}
e(K) = \frac{1}{\pi \sqrt{2}} \sum_i ( - Y_{O(i) x(i)} + Y_{O(i) y(i)} + Y_{x(i) y(i)})^2\sigma_{i, (2)}
\end{equation}
the constraint from topology is then 
\begin{equation}
\label{constraint}
\int W(e(K)) = 2.
\end{equation}
 At the first step, an edge is picked, $Y_{ij}$. Next, we find each two-simplex of which it is an edge, making the list of two simplices $k \in I$ if $[ij] < \sigma_{k, (2)}$. In turn we take all edges which are boundary of these simplices apart from $[ij]$ itself. The value of $Y$ is updated for each of these edges such that the Euler class is left invariant each such edge is update with the same value $\delta Y'$, after the two updates $Y \mapsto Y(\delta Y, \delta Y')$. The invariance of the Euler class means the following integral equation
\begin{equation}
\int W(e(K)(Y) ) =  \int W(e(K) (Y(\delta Y, \delta Y'))).
\end{equation}
The update $\delta Y'$ is a topological correction to the metric update $\delta Y$. 

The advantage of the present approach, with its systematic discretisation of differential geometry is that actions (at least bare actions) for matter fields can be readily constructed as 
\begin{equation}
S_M = \int \bar{W} (\bar{\Phi}^K) \wedge \bar{W} ( \bigstar^{\theta} D^K \Phi^K)
\end{equation}
where $\Phi^K$ is some physical field and $D^K$ the differential operator. In the case of fermionic fields one must keep in mind the doubling problem and so one may consider including a Wilson type term using the Laplacian $\triangle^K$.

As the third application considered, we are interested in the discretisation of Maxwell's equations (see e.g \cite{tex}). In the gauge theory formalism the field strength is
\begin{equation}
F = dA,
\end{equation}
which in terms of the physical fields $E$ and $B$ is
\begin{equation}
F = B + E \wedge dt.
\end{equation}
Then, Maxwell's equations are simply
\begin{eqnarray}
d F &=& 0 \\
\delta F &=& j.
\end{eqnarray}
In the 4D formalism, the interior product is used to express the boundary conditions which may be written as \cite{warnick}
\begin{eqnarray}
i_n(n \wedge (F_2 - F_1)) &=& 0 \\
i_n ( n \wedge ( \star F_2 - \star F_1)) &=& j_s 
\end{eqnarray}
where the subscript $s$ refers to surface
\begin{equation}
j_s = \rho_s - J_s \wedge dt.
\end{equation}
and the vector $n = \frac{df}{|df|}$  where $f$ is a function such that $df$ is orthogonal to the boundary. The present formulation may be used to access such situation in the discrete setting. A full implementation is needed to confirm this point. 

On the conceptual side, we note that in the literature a distinction is made between forms which are represented by chains and twisted forms represented in the dual complex. This has much to do with the Hodge star operator, if one look at the $(3+1)$D formulation, i.e the discretisation of the spatial part of the Maxwell's equations, we have the metric dependent part of the Maxwell's equations given by
\begin{eqnarray}
D &=& \epsilon \star E, \\
B &=& \mu \star H.
\end{eqnarray}
The fields $E$ and $H$ are arguably forms and the fields $D$ and $B$ are twisted forms. Simplicially, one argues that twisted forms have an outer orientation, in fact this is just the situation of a volume form or density in the continuum. In the continuum, one may introduce a pairing to represent such twisted forms as $(\alpha, \Omega)$  where $\alpha$ is a form and $\Omega$ is a volume element \cite{burke}. This pairing is interesting here since it can be identified with the pairing $\times$ central to this paper, so that the latter appears also as a discretisation of the pairing of Burke in the context of the electromagnetic theory.

\section{Conclusion}
A discretisation scheme has been proposed in which some key aspects of differential 
geometry relevant to physics are captured. A product space of chains introduced here and the 
extension of the Whitney map for the product space played a central role 
in the construction. Furthermore, the vielbein formulation of metric is 
consistently discretized in a smooth way and topological invariants are 
used to normalize some parameters in the metric. For comparison we 
established a correspondence, but no identification with the 
non-commutative geometry of graphs which we argued is incomplete as to capturing 
continuum like metric data. Three applications to physics are sketched, and are the focus of ongoing work.
\subsection*{Acknowledgments}
The work of V. de B was supported by the Japan Society for the Promotion of Science.

\end{document}